\documentclass{article}
\usepackage[utf8]{inputenc}
\usepackage[english]{babel}
\usepackage[letterpaper, vmargin = 1in, inner = 1.5in, outer = 1in]{geometry}

\usepackage{graphicx}

\usepackage{natbib}

\usepackage{mathtools}
\usepackage{xcolor}
\definecolor{greytext}{gray}{0.5}

\usepackage{appendix}
\usepackage[byname]{smartref}
\usepackage{enumitem}

\usepackage[allbordercolors={0.8 0.8 0.8}]{hyperref}

\usepackage{txfonts}
\usepackage{tocloft}
\usepackage{longtable}
\usepackage{algorithm}
\usepackage{algpseudocode}
\usepackage{etoolbox}

\usepackage{amsfonts,amssymb}

\usepackage{setspace}
\usepackage{epsfig}
\usepackage[nottoc]{tocbibind}

\usepackage{diagbox}
\usepackage{rotating}

\usepackage{cancel}
\usepackage{MnSymbol,bbding,pifont}

\usepackage{setspace}
\onehalfspace

\newcommand{\scenario}[1]{\textbf{Scenario~#1:}}

\newcommand*\rfrac[2]{{}^{#1}\!/_{#2}}

\usepackage{booktabs}
\usepackage{multirow}
\usepackage{caption}
\usepackage{lscape}
\usepackage{subfig}
\usepackage{setspace}
\usepackage{caption}
\usepackage{amsfonts}
\usepackage{tikz}
\usepackage{verbatim}
\usetikzlibrary{arrows.meta, shapes.geometric, calc, shadows, arrows}

\usepackage[group-separator={,},group-minimum-digits={3}]{siunitx} 

\usepackage{xr}
\makeatletter

\newcommand*{\addFileDependency}[1]{
\typeout{(#1)}
\@addtofilelist{#1}
\IfFileExists{#1}{}{\typeout{No file #1.}}
}\makeatother

\newcommand*{\myexternaldocument}[1]{%
\externaldocument{#1}%
\addFileDependency{#1.tex}%
\addFileDependency{#1.aux}%
}

\myexternaldocument{supplementary}

\makeatletter
\title{Model-based Clustering of Multi-Dimensional Zero-Inflated Counts via the EM Algorithm}
\author{
Zahra AghahosseinaliShirazi\textsuperscript{a}\thanks{ Email: zaghahos@uwo.ca}\,, Pedro A. Rangel\textsuperscript{a}\thanks{ Email:  pedroar.est@gmail.com} \, and Camila P. E. de Souza\textsuperscript{a}\thanks{Email: camila.souza@uwo.ca}\\
\textsuperscript{a}The University of Western Ontario, ON, Canada
}

\date{}

\begin{document}

\maketitle

\begin{abstract}

\noindent Zero-inflated count data arise in various fields, including health, biology, economics, and the social sciences. These data are often modelled using probabilistic distributions such as zero-inflated Poisson (ZIP), zero-inflated negative binomial (ZINB), or zero-inflated binomial (ZIB). To account for heterogeneity in the data, it is often useful to cluster observations into groups that may explain underlying differences in the data-generating process. This paper focuses on model-based clustering for zero-inflated counts when observations are structured in a matrix form rather than a vector.
We propose a clustering framework based on mixtures of ZIP or ZINB distributions, with both the count and zero components depending on cluster assignments. Our approach incorporates covariates through a log-linear structure for the mean parameter and includes a size factor to adjust for differences in total sampling or exposure. Model parameters and cluster assignments are estimated via the Expectation-Maximization (EM) algorithm. We assess the performance of our proposed methodology through simulation studies evaluating clustering accuracy and estimator properties, followed by applications to publicly available datasets.

\end{abstract}

\section{Introduction}

Count data often exhibit an excess of zeros due to various underlying mechanisms, and two common modelling approaches are hurdle models and zero-inflated models \citep{zeileis2008regression,faraway2016extending}. Hurdle models assume a two-step process where one mechanism determines whether an observation is zero or positive, while another governs the positive counts. In contrast, zero-inflated models account for excess zeros by including a latent process that generates additional zeros beyond those expected from a standard count distribution. This study focuses on zero-inflated Poisson (ZIP) and zero-inflated negative binomial (ZINB) models, which are widely utilized in fields such as health, biology, economics, and psychology, extending classical count distributions to accommodate excess zeros \citep{Lambert, Farewell, Hall, Hu}. In the presence of covariates, regression frameworks based on ZIP or ZINB have been employed to analyze diverse datasets. For instance, \cite{lyashevska2016mapping} and \cite{Pilosof2021} used ZIP and ZINB regression models, respectively, to model species abundance. Additionally, \cite{Xiaonan2020} introduced a ZIP regression model with random intercepts to analyze adult physical activity levels.

To address potential data heterogeneity, some studies extend zero-inflated models by allowing the count distribution component to be a mixture of distributions, focusing on the mixture itself rather than clustering the data. For example, \cite{ZIPMIXREG2014}, \cite{RNA_seq_mixture_Poisson}, and \cite{ZIPRM} proposed ZIP regression models where the Poisson component is modelled as a mixture of Poisson distributions. These studies estimate model parameters using the Expectation-Maximization (EM) algorithm \citep{Dempster1977, Mclachlan2007}. As applications of their approach, \cite{ZIPMIXREG2014}, \cite{RNA_seq_mixture_Poisson}, and \cite{ZIPRM} employed their methods to analyze counts of dental caries in adolescents, RNA sequencing counts, and counts of adverse events during gastrointestinal endoscopy, respectively. In a Bayesian framework, \cite{Morgan} developed a ZIP regression model with a finite mixture of Poisson distributions for the count component. Motivated by endophenotype schizophrenia family data, they incorporated random effects to account for correlations among observations within the same family.

Focusing on model-based clustering of zero-inflated count data, we note some methods developed in the context of single-cell RNA sequencing (scRNA-seq). \cite{ZINBMM} introduce a ZINB mixture model with cluster-specific mean and always-zero probabilities, while keeping dispersion parameters constant across clusters. To account for batch effects, their model allows the mean (rate) parameters to depend on a linear combination of covariates via a log link. Estimation is performed via the EM algorithm. Similarly, \cite{RZIMM} propose ZIP and ZINB mixture models, but with only the count component modelled as a mixture of distributions. Unlike \cite{ZINBMM}, they do not incorporate covariates, and model parameters are estimated via maximum likelihood using a majorization-maximization algorithm.

While previous work has explored mixtures in zero-inflated models, most studies focus on modelling the mixture without explicitly addressing clustering assignments. Additionally, existing approaches typically apply mixtures only to the Poisson or negative binomial component, often without incorporating covariates or a size factor. In this paper, we propose a model-based approach for clustering heterogeneous subjects based on multiple independent count observations, explicitly accounting for the zero-inflated nature of the data. Our data structure, which considers multiple observations per subject, is inspired by scRNA-seq data, where genes provide repeated measurements across cells. We model the data as a mixture of ZIP or ZINB distributions, allowing both the count and zero components to depend on the cluster. Furthermore, we incorporate covariates—such as treatment or batch effects—through a log-linear structure for the mean (rate) parameter, including an offset size factor. This size factor adjusts for differences in the total amount of sampling or exposure, which influence the expected number of counts. For example, in scRNA-seq data, library size variation affects gene expression counts, while in other contexts, the size factor may account for differences in observation time or population size. Cluster assignments and estimates of model parameters are obtained via the EM algorithm. We evaluate clustering performance and the statistical properties of our estimators, including bias and variance, through extensive simulation studies under various scenarios. Finally, we demonstrate the practical utility of our model-based clustering approach by applying it to two publicly available scRNA-seq datasets \citep{inDrop, Tissues}. Our methodology is implemented in R, with the code available at \url{https://github.com/desouzalab/em-mzip}.

\indent This paper is organized as follows. Section \ref{sec:method} describes our model-based methodology for clustering zero-inflated counts. Section \ref{Results} presents simulation results under various scenarios. Results from our real data applications are shown in Section \ref{ch:data_analysis}. Finally, the conclusion and final remarks are discussed in Section \ref{ch:conclusion}.

\section{Method} \label{sec:method}
Our model-based clustering approach assumes that the zero-inflated count data arise from a mixture of zero-inflated Poisson (ZIP) or zero-inflated negative binomial (ZINB) distributions. Section \ref{proposed_mix_zip} presents the proposed mixture model and EM algorithm for zero-inflated Poisson counts, considering the cases without and with covariates. In Section \ref{proposed_mix_ZINB}, we describe our proposed methodology for the mixture of ZINB distributions, also covering cases without and with covariates.

\subsection{The proposed mixture model for ZIP counts}\label{proposed_mix_zip}
  
Let \( Y_{ng} \) be a random variable representing the zero-inflated count for observation \( g \) within subject \( n \), where \( g = 1, \dots, G \) and \( n = 1, \dots, N \). The variable \( Y_{ng} \) takes values in \( \{0,1,2,3,\dots\} \). Thus, the observed data can be expressed in the following matrix format:

\begin{equation} \label{Y_matrix_CH4}
\mathbf{y}= \begin{pmatrix}
y_{11} & y_{12} & \cdots& y_{1G}\\
y_{21} & y_{22} & \cdots &y_{2G}\\
\vdots &\vdots &\ddots &\vdots \\
 y_{N1}&y_{N2} &\cdots &y_{NG}
\end{pmatrix}.
\end{equation}

Suppose that there are $K \ll N$ clusters of subjects and let $\boldsymbol{Z}=\{Z_{11},\dots,Z_{NK}\}$
be the set of latent random variables indicating the true subject cluster assignments, that is:
\begin{align}\label{Z_nk}
Z_{nk} = \begin{dcases*}
    1, & if subject $n$ belongs to cluster $k$, \\
    0, & otherwise.
\end{dcases*}
\end{align}
\noindent with $\sum_{k=1}^K Z_{nk}=1$, $P(Z_{nk}=1)=\pi_k$ for $n=1,\dots,N$, $k=1,\dots, K$, and $\sum_{k=1}^{K}\pi_k=1$.
We assume that given $Z_{nk}=1$, $Y_{n1},\ldots,Y_{nG}$ are independent and follow a ZIP distribution with parameters that depend on cluster $k$. Let $\boldsymbol{\theta}=\{\boldsymbol{\theta}_1,\dots,\boldsymbol{\theta}_K\}$ be the set of all models parameters with $\boldsymbol{\theta}_k=\{\pi_k,\phi_k,\boldsymbol{\lambda}_k\}$, and $\boldsymbol{\lambda_k}=\{\lambda_{1k},\dots,\lambda_{Gk}\}$.
Thus, we can write the probability mass function (pmf) for each subject as the following mixture of zero-inflated Poisson (ZIP) distributions: 
$$p(\mathbf{y}_n\,|\,\boldsymbol{\theta})=\sum_{k=1}^K \pi_k p(\mathbf{y}_n \,|\,\boldsymbol{\theta}_k)=\sum_{k=1}^K \pi_k \prod_{g=1}^G p(y_{ng}\,|\,\lambda_{gk},\phi_k),$$ 
where 
\begin{align}\label{P_Yng}
 p(y_{ng}\,|\,\lambda_{gk},\phi_k) = \begin{dcases*}
    \phi_k+(1-\phi_k)e^{-\lambda_{gk}}, & if $y_{ng}=0$, \\
    (1-\phi_k)\frac{e^{-\lambda_{gk}} \lambda_{gk}^{y_{ng}}}{y_{ng}!}, & if $y_{ng}=1,2,3, \dots$,
\end{dcases*}
\end{align}
in which $\lambda_{gk}$ is the Poisson rate parameter and $\phi_k$ is the probability of always (or perfect) zero. By assuming independence across subjects, the observed-data log-likelihood based on all subjects is given by:
\begin{equation}\label{obs_ZI1}
    \ell(\boldsymbol{\theta}\,|\,\boldsymbol{y})=\sum_{n=1}^N \log \left[\sum_{k=1}^K \pi_k \prod_{g=1}^G p(y_{ng}\,|\,\lambda_{gk},\phi_k)\right].
\end{equation}

The goal is to find the parameter estimates that maximize $\ell(\boldsymbol{\theta}\,|\,\boldsymbol{y})$ in \eqref{obs_ZI1}, but we cannot do it analytically with a closed formula.
To tackle this problem, we develop an EM algorithm to iteratively find the parameter estimates.  
To obtain these estimates using the EM framework, we consider the true latent cluster assignments $Z_{nk}$ for $n=1,\dots,N$ and $k=1,\dots,K$ as in (\ref{Z_nk}). In addition, we introduce another set of hidden variables $\boldsymbol{U}=\{U_{11},\ldots,U_{NG}\}$, where each $U_{ng}$, for $n=1,\ldots,N$ and $g=1,\ldots,G$, is defined as follows:
\[ U_{ng}= \begin{dcases*}
    1, & if $y_{ng}$ is from the perfect zero state, \\
    0, & if $y_{ng}$ is from the Poisson state.
\end{dcases*} \]
This latent indicator variable is drawn from a Bernoulli distribution that depends on the cluster $k$, $U_{ng} \mid Z_{nk} = 1 \sim$ Bernoulli($\phi_k$), with probability of success (for this case, probability of always zero) $\phi_k$, defined as
$\phi_k = P(U_{ng} = 1 | Z_{nk} = 1)$.

In the E-step of the EM algorithm to maximize the observed-data log-likelihood, first we obtain the conditional expectation of the complete-data log-likelihood, and maximize it given the observed data and current parameter estimates is obtained.
Then, in the M-step, the expectation from the E-step is  maximized with respect to each parameter of interest.
These two steps are executed iteratively until convergence. Now, considering the observed counts and the introduced latent variables, the complete-data log-likelihood can be written as:
\begin{align*}
\ell(\boldsymbol{\theta}\,|\,\boldsymbol{y},\boldsymbol{u},\boldsymbol{z}) =&
\sum_{n=1}^N \sum_{k=1}^K z_{nk}\log \pi_k
+ \sum_{n=1}^N \sum_{k=1}^K \sum_{g=1}^G \Big(z_{nk}u_{ng}\log \phi_k+z_{nk}(1-u_{ng})\log(1-\phi_k)\Big) \\
&+ \sum_{n=1}^N \sum_{k=1}^K \sum_{g=1}^G z_{nk}(1-u_{ng})\log\left(\frac{e^{-\lambda_{gk}}\lambda_{gk}^{y_{ng}}}{y_{ng}!}\right).
\end{align*}
In what follows, we present the E and M steps of our proposed EM algorithm for the case without covariates (Section \ref{ZIP_NOX}) and the case with covariates (Section \ref{ZIP_X}).
\subsubsection{EM for the ZIP mixture model without covariates}\label{ZIP_NOX}
\textbf{E-Step:} We first write the conditional expectation of the complete-data log-likelihood given the current estimates of the parameters, 
$\boldsymbol{\theta}^{(t)}=\{\boldsymbol{\lambda}^{(t)},\,\boldsymbol{\phi}^{(t)},\,\boldsymbol{\pi}^{(t)}\}$, and the observed data, $\boldsymbol{y}$: 
\begin{align}
\label{ZIP_M1}
Q(\boldsymbol{\theta};\boldsymbol{\theta}^{(t)}) =&
E\Big[\ell(\boldsymbol{\theta}\,|\,\boldsymbol{y},\boldsymbol{u},\boldsymbol{z})\, |\,\boldsymbol{y},\boldsymbol{\theta}^{(t)}\Big]  \nonumber \\
=& \sum_{n=1}^N \sum_{k=1}^K E\Big[Z_{nk}\,|\,\boldsymbol{y},\boldsymbol{\theta}^{(t)}\Big] \log \pi_k \nonumber \\
&+ \sum_{n=1}^N \sum_{k=1}^K \sum_{g=1}^G \Bigg(E\Big[Z_{nk}U_{ng} \,|\, \boldsymbol{y},\boldsymbol{\theta}^{(t)}\Big] \log \phi_k+ 
E\Big[Z_{nk}(1-U_{ng})\,|\, \boldsymbol{y},\boldsymbol{\theta}^{(t)}\Big]\log(1-\phi_k) \Bigg) \nonumber \\
&+ \sum_{n=1}^N \sum_{k=1}^K \sum_{g=1}^G E\Big[Z_{nk}(1-U_{ng})\,|\,\boldsymbol{y},\boldsymbol{\theta}^{(t)}\Big] \log\Bigg(\frac{e^{-\lambda_{gk}}\lambda_{gk}^{y_{ng}}}{y_{ng!}}\Bigg)
\end{align}
Then, we can show that the expectations in (\ref{ZIP_M1}) are 
as follows:
\begin{align} \label{Z_nk1}
\hat{Z}_{nk}^{(t)}&=E\Big[Z_{nk}\,|\,\boldsymbol{y}\,,\,\boldsymbol{\theta}^{(t)}\Big] \nonumber \\ 
&=E\Big[Z_{nk}\,|\,\boldsymbol{y}_n\, ,\,\boldsymbol{\theta}^{(t)}\Big] \nonumber \\
&=p\Big(Z_{nk}=1\,|\,\boldsymbol{y}_n\,,\,\boldsymbol{\theta}^{(t)}\Big) \nonumber\\
&=\frac{\pi_k^{(t)}\prod_{g=1}^G p\Big(y_{ng}\,|\,\lambda_{kg}^{(t)},\phi_k^{(t)}\Big)}{\sum_{j=1}^K \pi_{j}^{(t)}\prod_{g=1}^G p\Big(y_{ng}\,|\,\lambda_{jg}^{(t)},\phi_j^{(t)}\Big)}\mbox{,}
\end{align}
and
\begin{align*}
E\big[ Z_{nk}U_{ng}\, \big|\, \boldsymbol{y},\boldsymbol{\theta}^{(t)}\big]
&= p(Z_{nk}=1,U_{ng}=1\,\big|\,\boldsymbol{y},\boldsymbol{\theta}^{(t)}) \\
&= p(U_{ng}=1\,\big|\,Z_{nk}=1,y_{ng},\boldsymbol{\theta}^{(t)})\,\times\, p(Z_{nk}=1\,\big|\,\boldsymbol{y}_n,\boldsymbol{\theta}^{(t)})=\hat{U}_{ngk}^{(t)}\hat{Z}_{nk}^{(t)},
\end{align*}
\noindent where $\hat{Z}_{nk}^{(t)}$ is as in (\ref{Z_nk1}) and $\hat{U}_{ngk}^{(t)} = p(U_{ng}=1\,\big|\,Z_{nk}=1,y_{ng},\boldsymbol{\theta}^{(t)}) $ is given by: 

\begin{equation}\label{U_ng}
 \hat{U}_{ngk}^{(t)}=
\begin{dcases*}
\frac{\phi_k^{(t)}}{\Big(\phi_k^{(t)}+(1-\phi_k^{(t)})e^{-\lambda_{gk}^{(t)}}\Big)}, & if $y_{ng}=0$, \\
0, & if $y_{ng}=1,2,\dotsc\ $.    
\end{dcases*}
\end{equation}

\noindent Note that $\hat{U}_{ngk}^{(t)}$ in (\ref{U_ng}) is separated into two cases because if $U_{ng}=1$, $y_{ng}$ can only be equal to zero, otherwise, if $y_{ng}$ takes a non-zero count value, it definitely arises from the Poisson state.

Using the calculated values $\hat{Z}_{nk}^{(t)}$ and $\hat{U}_{ngk}^{(t)}$ from Equations (\ref{Z_nk1}) and (\ref{U_ng}), we can rewrite $Q(\boldsymbol{\theta};\boldsymbol{\theta}^{(t)})$ in (\ref{ZIP_M1}) as:

\begin{align}\label{ZIP_Q1}
Q(\boldsymbol{\theta};\boldsymbol{\theta}^{(t)})&=Q_1(\boldsymbol{\pi};\boldsymbol{\pi}^{(t)})+Q_2(\boldsymbol{\phi};\boldsymbol{\phi}^{(t)})+Q_3(\boldsymbol{\lambda};\boldsymbol{\lambda}^{(t)}),
\end{align}
where 
\begin{align}
Q_1(\boldsymbol{\pi};\boldsymbol{\pi}^{(t)}) =& \sum_{n=1}^N \sum_{k=1}^K \hat{Z}_{nk}^{(t)} \log(\pi_k), \label{Q1_b} \\
Q_2(\boldsymbol{\phi};\boldsymbol{\phi}^{(t)}) =& \sum_{n=1}^N \sum_{k=1}^K \sum_{g=1}^G \Big[\hat{Z}_{nk}^{(t)} \hat{U}_{ngk}^{(t)}\log(\phi_k)+\hat{Z}_{nk}^{(t)} (1-\hat{U}_{ngk}^{(t)})\log(1-\phi_k)\Big], \label{Q2_b} \\
\intertext{and}
Q_3(\boldsymbol{\lambda};\boldsymbol{\lambda}^{(t)}) =& \sum_{n=1}^N \sum_{k=1}^K \sum_{g=1}^G \Big\{\hat{Z}_{nk}^{(t)} (1-\hat{U}_{ngk}^{(t)})\Big[-\lambda_{gk}+y_{ng}\log\lambda_{gk}-\log y_{ng}! \Big]\Big\}. \nonumber
\end{align}

\textbf{M-step:} In this step, we maximize $Q(\boldsymbol{\theta};\boldsymbol{\theta}^{(t)})$ in (\ref{ZIP_Q1}) with respect to $\boldsymbol{\theta}$, obtaining the following updated parameter estimates: 
\begin{align}
\pi_k^{(t+1)} =& \frac{\sum_{n=1}^N \hat{Z}_{nk}^{(t)}}{N}, \label{P_1} \\
\phi_k^{(t+1)} =& \frac{\sum_{n=1}^N \sum_{g=1}^G\hat{Z}_{nk}^{(t)} \hat{U}_{ngk}^{(t)}} {G \sum_{n=1}^N \hat{Z}_{nk}^{(t)}}, \text{ and} \label{Phi_1} \\
\lambda_{gk}^{(t+1)} =& \frac{\sum_{n=1}^N  \hat{Z}_{nk}^{(t)}(1-\hat{U}_{ngk}^{(t)}) y_{ng}}{\sum_{n=1}^N  \hat{Z}_{nk}^{(t)}(1-\hat{U}_{ngk}^{(t)})}. \label{lambda_1}
\end{align}

Note that the conditional expected value of each $Z_{nk}$ in Equation (\ref{Z_nk1}), obtained at the last (upon convergence) iteration $t^*$, is used to infer the cluster assignment of each subject. Thus, we obtain the decision that subject $n$ belongs to cluster $k$ if that cluster is the one with the highest expected value (highest probability); that is:   
\begin{equation}
\hat{Z}_{nk}=
\begin{dcases*}
1, & if $\hat{Z}_{nk}^{(t^*)}=\underset{j \in \{1,\dots,K\}}{\mbox{max}}{\hat{Z}_{nj}^{(t^*)}}$, \\
0, & otherwise.
\end{dcases*}
\label{eq:determine_cluster}
\end{equation}

Algorithm \ref{alg::ZIP_No_X} in the Appendix summarizes the EM algorithm for the ZIP mixture model without covariates.


\subsubsection{EM for the ZIP mixture model with covariates}\label{ZIP_X}

In this case, we assume that the Poisson rate parameters depend on a linear combination of covariates via a log link function as follows:
\begin{equation}\label{full_lambda}
    \log\lambda_{ngk}=\log T_n+\rho_{gk}+\beta_{0g}+\sum_{p=1}^P \beta_{pg} x_{np},
\end{equation}
for $n=1,\dotsc,N$, $g=1,\dotsc,G$, $k=1,\dotsc,K$, and $p=1,\dotsc,P$,
where $T_n$ is a fixed size factor (also known as a Poisson offset) for subject $n$,
$\rho_{gk}$ is the fixed effect of cluster $k$ on observation $g$,
$\beta_{0g}$ is a baseline expression for observation $g$,
$x_{n1},\dotsc,x_{nP}$ are $P$ known covariates for subject $n$ (e.g., batch or treatment effects), and $\beta_{1g},\dotsc,\beta_{Pg}$ their corresponding unknown coefficients on observation $g$. 

This model with covariates can also be called a mixture of generalized ZIP regression models. We again use the EM algorithm to find the estimated parameters and inferred cluster assignments. Therefore, considering the complete-data log-likelihood as
\begin{align*}
\ell(\boldsymbol{\theta}\,|\,\boldsymbol{y},\boldsymbol{x},\boldsymbol{z},\boldsymbol{u})
= \sum_{n=1}^N \sum_{g=1}^G \sum_{k=1}^K \Big[
& z_{nk}  \log\pi_k+z_{nk} u_{ng}\log\phi_k+z_{nk}(1-u_{ng})\,\log(1-\phi_k) \\
&+ \;z_{nk}(1-u_{ng}) \,\log(p(y_{ng} \mid \rho_{gk},\beta_{0g},\beta_{pg}))\Big],  
\end{align*}
where $\boldsymbol{\theta}=(\boldsymbol{\pi},\boldsymbol{\phi},\boldsymbol{\rho},\boldsymbol{\beta}_{0},\boldsymbol{\beta})$, with 
$\boldsymbol{\pi}=(\pi_1,\dots,\pi_k)^T$,
$\boldsymbol{\phi}=(\phi_1,\dots,\phi_k)\,^T$,
$\boldsymbol{\rho}=(\rho_{11},\dots,\rho_{G1},\dots,\rho_{1K},\dots,\rho_{GK})^T$,
$\boldsymbol{\beta}_0=(\beta_{01},\dots,\beta_{0G})^T$
and
$\boldsymbol{\beta}=(\beta_{11},\dots,\beta_{P1},\dots,\beta_{1G},\dots,\beta_{PG})^T$. \\

\paragraph{E-Step:} Similarly to Section \ref{ZIP_NOX}, first in the E-step, we compute the conditional expectation of the complete-data log-likelihood given the observed data and the current parameter estimates as follows:
\begin{align} \label{Q_C2}
Q(\boldsymbol{\theta};\boldsymbol{\theta}^{(t)}) =&
E\Big[\ell(\boldsymbol{\theta}\,|\, \boldsymbol{y},\,\boldsymbol{x},\,\boldsymbol{z},\,\boldsymbol{u}\,)\,|\, \boldsymbol{y},\,\boldsymbol{x},\,\boldsymbol{\theta}^{(t)}\, 
\Big] \nonumber \\ 
=& \sum_{n=1}^N \sum_{k=1}^K E\Big[Z_{nk}\,|\,\boldsymbol{y},\boldsymbol{x},\boldsymbol{\theta}^{(t)}\Big] \log(\pi_k) \nonumber \\
&+ \sum_{n=1}^N \sum_{g=1}^G \sum_{k=1}^K \Big(E\Big[Z_{nk}U_{ng}\,|\,\boldsymbol{y},\boldsymbol{x},\boldsymbol{\theta}^{(t)}\Big] \log(\phi_k)+E\Big[Z_{nk}(1-U_{ng})\,|\,\boldsymbol{y},\boldsymbol{x},\boldsymbol{\theta}^{(t)}\Big]\log(1-\phi_k)\Big) \nonumber \\
&+ \sum_{n=1}^N \sum_{g=1}^G \sum_{k=1}^K  E\Big[Z_{nk}(1-U_{ng})\,|\,\boldsymbol{y},\boldsymbol{x},\boldsymbol{\theta}^{(t)}\Big] \nonumber \\
& \times \left[-\exp\left\{\log T_n +\beta_{0g}+\rho_{gk}+\sum_{p=1}^P\,\beta_{pg}x_{np}\right\} + y_{ng}\, \left\{\log T_n +\beta_{0g}+\rho_{gk}+\sum_{p=1}^P\,\beta_{pg}x_{np}\right\}-\log\, y_{ng}! \right]
\end{align} 
Similarly to Equations (\ref{Z_nk1}) and (\ref{U_ng}) in Section \ref{ZIP_NOX}, we calculate $\hat{Z}_{nk}^{(t)}$ and $\hat{U}_{ngk}^{(t)}$ as follows:
\begin{equation}\label{Z_nk2}
\hat{Z}_{nk}^{(t)}=E\Big[Z_{nk}\,|\,\boldsymbol{y},\boldsymbol{x},\boldsymbol{\theta}^{(t)} \Big]=\frac{\pi_k^{(t)}\prod_{g=1}^G p(y_{ng}\,|\,\phi_k^{(t)},\rho_{gk}^{(t)},\beta_{0g}^{(t)},\beta_{pg}^{(t)})}{\sum_{k=1}^K\pi_k^{(t)}\prod_{g=1}^G p(y_{ng}\,|\,\phi_k^{(t)},\rho_{gk}^{(t)},\beta_{0g}^{(t)},\beta_{pg}^{(t)})}
\end{equation}
\begin{equation}\label{U_ng2}
 \hat{U}_{ngk}^{(t)}=p\Big(U_{ng}=1|Z_{nk}=1,\boldsymbol{x},y_{ng},\boldsymbol{\theta}^{(t)}\Big)=
\begin{dcases*}
    \frac{\phi_k^{(t)}}{\Big(\phi_k^{(t)}+(1-\phi_k^{(t)})e^{-\lambda_{ngk}^{(t)}}\Big)}, & if $y_{ng}=0$, \\
    0, & if $y_{ng}=1,2,\dotsc,$
\end{dcases*}
\end{equation}
where $\lambda_{ngk}^{(t)}=\exp\Big\{\log T_n +\rho_{gk}^{(t)}+\beta_{0g}^{(t)}+\sum_{p=1}^P \beta_{pg}^{(t)} x_{np}\Big\}$.

By using the expected values $\hat{Z}_{nk}^{(t)}$ and $\hat{U}_{ngk}^{(t)}$ from Equations (\ref{Z_nk2}) and (\ref{U_ng2}), we can rewrite $Q(\boldsymbol{\theta};\boldsymbol{\theta}^{(t)})$ as follows:
\begin{align*}
Q(\boldsymbol{\theta}\,;\,\boldsymbol{\theta}^{(t)})&=Q_1(\boldsymbol{\pi}\,;\,\boldsymbol{\pi}^{(t)})+Q_2(\boldsymbol{\phi}\,;\,\boldsymbol{\phi}^{(t)})+Q_3\Big((\boldsymbol{\beta}_{0},\,\boldsymbol{\rho},\,\boldsymbol{\beta})\,;\,(\boldsymbol{\beta}_{0}^{(t)},\,\boldsymbol{\rho}^{(t)},\,\boldsymbol{\beta}^{(t)}\,)\Big)
\end{align*}
where the form of $Q_1(\boldsymbol{\pi}\,;\,\boldsymbol{\pi}^{(t)})$ and $Q_2(\boldsymbol{\phi}\,;\,\boldsymbol{\phi}^{(t)})$ remain as in (\ref{Q1_b}) and (\ref{Q2_b}), respectively, and 
\begin{align} \label{FullQ_3}
&  Q_3\Big((\boldsymbol{\beta}_{0},\,\boldsymbol{\rho},\,\boldsymbol{\beta}\,);\,(\boldsymbol{\beta}_{0}^{(t)},\boldsymbol{\rho}^{(t)},\,\boldsymbol{\beta}^{(t)})\Big)
= \sum_{n=1}^N \sum_{g=1}^G \sum_{k=1}^K \hat{Z}_{nk}^{(t)} (1-\hat{U}_{ngk}^{(t)}) \times \nonumber \\
&\left[-\exp\left\{\log(T_n)+\beta_{0g}+\rho_{gk}+\sum_{p=1}^P\,\beta_{pg}x_{np}\right\} + y_{ng} \left\{ \log(T_n)+\beta_{0g}+\rho_{gk}+\sum_{p=1}^P\,\beta_{pg}x_{np} \right\} - \log\, y_{ng}! \right]
\end{align}

\noindent \textbf{M-Step:} In the M-Step, through differentiating $Q(\boldsymbol{\theta}\,;\,\boldsymbol{\theta}^{(t)})$
with respect to each parameter, we can find the updated estimates. We obtain $\pi_k^{(t+1)}$ and $\phi_k^{(t+1)}$ in closed form as in (\ref{P_1}) and (\ref{Phi_1}), respectively, in Section \ref{ZIP_NOX}. Next, we find the updated estimates $\beta_{0g}^{(t+1)}$, $\rho_{gk}^{(t+1)}$, and $\beta_{pg}^{(t+1)}$ as the values that maximize $Q_3$ in (\ref{FullQ_3}) numerically by applying the Fisher scoring algorithm (a form of Newton--Raphson method) within the M-step.  Algorithm \ref{alg::ZIP_X} in the Appendix presents a summary of the EM algorithm steps for the ZIP mixture model with covariates.

A simpler model than in (\ref{full_lambda}) can be considered when there are no covariates, but one wants to still include an offset of size factor $T_n$. Thus, we can model $\lambda_{ngk}$ as 
\begin{equation}\label{reduced_lambda}
    \log\lambda_{ngk} = \log T_n + \rho_{gk}+\beta_{0g}.
\end{equation}

In this case, only a few modifications are required in the EM algorithm. In the E-step, $Q_1(\boldsymbol{\pi}\,;\,\boldsymbol{\pi}^{(t)})$ and $Q_2(\boldsymbol{\phi}\,;\,\boldsymbol{\phi}^{(t)})$ remain as in (\ref{Q1_b}) and (\ref{Q2_b}), respectively, but $Q_3$ now becomes:
\begin{align*}
\footnotesize 
Q_3\big((\boldsymbol{\beta}_{0},\,\boldsymbol{\rho})\,;\,(\boldsymbol{\beta}_{0}^{(t)},\,\boldsymbol{\rho}^{(t)})\big)=\sum_{n=1}^N \sum_{g=1}^G \sum_{k=1}^K \hat{Z}_{nk}^{(t)} (1-\hat{U}_{ngk}^{(t)})\, \times \left[-\exp \left\{\log(T_n)+\beta_{0g}+\rho_{gk} \right\} + y_{ng} \left\{\log(T_n)+\beta_{0g}+\rho_{gk} \right\}-\log\, y_{ng}! \right].
\end{align*}
Therefore, in the M-step, the updates $\pi_k^{(t+1)}$ and $\phi_k^{(t+1)}$ are as in (\ref{P_1}) and (\ref{Phi_1}), respectively, and the updates $\rho_{gk}^{(t+1)}$ and $\beta_{0g}^{(t+1)}$ can also be obtained via the Fisher scoring algorithm within the M-step. 

In our code implementation of the EM algorithm, to avoid identifiability issues when estimating the parameters, we assume that $\beta_{gk}=\beta_{0g}+\rho_{gk}$ with the restriction that $\sum_{k=1}^K \rho_{gk}=0$. This assumption and restriction imply that $\beta_{0g} = \sum_{k=1}^K \beta_{gk}/K$. So, we fit our model considering $\log \lambda_{ngk} = \log (T_n) + \beta_{gk}$ (or  $\log \lambda_{ngk} = \log (T_n) + \beta_{gk} + \sum_{p=1}^P \beta_{pg} x_{np}$), and after we obtain $\beta_{gk}^{(t+1)}$ at each EM iteration, we find the updates for $\beta_{0g}$ and $\rho_{gk}$ as follows:
 $$\beta_{0g}^{(t+1)}=\frac{\sum_{k=1}^K\beta_{gk}^{(t+1)}}{K}, \;\; \mbox{and} \; \;\rho_{gk}^{(t+1)}=\beta_{gk}^{(t+1)}- \; \beta_{0g}^{(t+1)}.$$

\subsection{The proposed mixture model for ZINB counts}\label{proposed_mix_ZINB}
In this section, our proposed clustering approach is based on a mixture model of zero-inflated negative binomial distributions. In this case, given $Z_{nk}=1$, that is, given that subject $n$ belongs to cluster $k$, we assume that observations are independent and follow a zero-inflated negative binomial (ZINB) distribution with parameters that depend on cluster $k$. Let $\boldsymbol{\mu}_k=(\mu_{11},\dots,\mu_{GK})^T$ and $\boldsymbol{\theta}_k=\{\pi_k,\phi_k,\alpha_k,\boldsymbol{\mu}_k\}$, where $\pi_k$ is the cluster assignment probability, defined as $P(Z_{nk} =1)=\pi_k$, $\phi_k$ is the zero-inflation proportion (or probability of always zero), $\mu_{gk}$ is the rate parameter and $\alpha_k $  is the dispersion parameter which is the inverse of size ($\nu_k$) parameter $(\alpha_k=\rfrac{1}{\nu_k})$.
Thus, we define $\boldsymbol{\theta}=\{\boldsymbol{\theta}_1, \dotsc, \boldsymbol{\theta}_K\}$ as the set containing all model parameters and write the probability mass function for each subject as a mixture of ZINB distributions as follows:

$$p(\boldsymbol{y}_n\,|\,\boldsymbol{\theta})=\sum_{k=1}^K \pi_k p(\mathbf{y}_n \,|\,\boldsymbol{\theta}_k)=\sum_{k=1}^K \pi_k \prod_{g=1}^G p(y_{ng}\,|\,\phi_k,\alpha_k,\mu_{gk}),$$ 
where  
\begin{align}\label{ZINB_2}
\scriptstyle p(y_{ng}\,|\,\phi_k,\alpha_k,\mu_{gk}) =
\begin{dcases*}
    \scriptstyle \phi_k,
    & \small if $y_{ng}$ belongs to the always zero state, \\
    \scriptstyle (1-\phi_k)\frac{\Gamma(y_{ng}+\frac{1}{\alpha_k})}{\Gamma(y_{ng}+1)\Gamma(\frac{1}{\alpha_k})}\left(\frac{1}{1+\alpha_k\mu_{gk}}\right)^{(\frac{1}{\alpha_k})}\left(1-\frac{1}{1+\alpha_k\mu_{gk}}\right)^{y_{ng}},
    & \small if $y_{ng}$ belongs to the NB state,
\end{dcases*}
\end{align}
\noindent for $n=1,\dots,N$, $g=1,\dots,G$, and $k=1,\dots,K$. The function $\Gamma(\cdot)$ is the gamma function. 

Thus, the observed-data log-likelihood across all subjects is given by: 
\begin{equation}\label{obs_ZINB1}
    \ell(\boldsymbol{\theta}\,|\,\boldsymbol{y})=\sum_{n=1}^N \log \left[\sum_{k=1}^K \pi_k \prod_{g=1}^G p(y_{ng}\,|\,\phi_k,\alpha_k,\mu_{gk})\right].
\end{equation}
Similarly to Section \ref{proposed_mix_zip}, our goal is to find the parameter estimates that maximize (\ref{obs_ZINB1}) iteratively via the EM algorithm.  To find the parameter estimates using the EM framework, we consider the latent cluster assignments 
$\boldsymbol{Z}=(Z_{11},\dots,Z_{NK})^T$ and the hidden Bernoulli variable $U_{ng}$ now defined as follows:
\[ U_{ng}=
\begin{dcases*}
1, & if $y_{ng}$ is from the perfect zero state,\\
0, & if $y_{ng}$ is from the Negative Binomial (NB) state,
\end{dcases*} \]
\noindent with $P(U_{ng}=1|Z_{nk} =1) = \phi_k$, for $n=1,\dots,N$ and $g=1,\dots,G$. 
Considering the observed counts and the latent random variables, we can write the completed-data log-likelihood as follows:

\begin{align*}
\ell(\boldsymbol{\theta}\,|\,\boldsymbol{y},\boldsymbol{u},\boldsymbol{z})
=& \sum_{n=1}^N \sum_{k=1}^K z_{nk}\log \pi_k. \\
&+ \sum_{n=1}^N \sum_{k=1}^K \sum_{g=1}^G \Big(z_{nk}u_{ng}\log (\phi_k)+z_{nk}(1-u_{ng})\log(1-\phi_k)\Big) \\
&+ \sum_{n=1}^N \sum_{k=1}^K \sum_{g=1}^G z_{nk}(1-u_{ng}) \log\left(
    \frac{\Gamma(y_{ng}+\frac{1}{\alpha_k})}{\Gamma(y_{ng}+1)\Gamma(\frac{1}{\alpha_k})}\left(\frac{1}{1+\alpha_k\mu_{gk}}\right)^{(\frac{1}{\alpha_k})}\left(1-\frac{1}{1+\alpha_k\mu_{gk}}\right)^{y_{ng}}
\right) \\
\end{align*}

In what follows, the E and M steps of our proposed EM algorithm for the ZINB mixture model without covariates (Section \ref{ZINB_NOX}) and with covariates (Section \ref{ZINB_X}) are presented.
\subsubsection{EM for the ZINB mixture model without covariates} \label{ZINB_NOX}
\vspace{0.2cm}
\textbf{E-Step:} Given the current estimates of the parameters $\boldsymbol{\theta}^{(t)}=(\boldsymbol{\alpha}^{(t)},\boldsymbol{\mu}^{(t)},\boldsymbol{\phi}^{(t)},\boldsymbol{\pi}^{(t)})$ and the observed data $\boldsymbol{y}$, we compute the conditional expectation of the complete-data log-likelihood as:
\begin{align}\label{ZINB_M1}
Q(\boldsymbol{\theta};\boldsymbol{\theta}^{(t)})
=& E\Big[\ell(\boldsymbol{\theta}\,|\,\boldsymbol{y},\boldsymbol{u},\boldsymbol{z})\,\big |\,\boldsymbol{y},\boldsymbol{\theta}^{(t)}\Big] \nonumber \\
=& \sum_{n=1}^N \sum_{k=1}^K E\Big[Z_{nk}\,\Big|\,\boldsymbol{y},\boldsymbol{\theta}^{(t)}\Big] \log(\pi_k) \nonumber \\
&+ \sum_{n=1}^N \sum_{k=1}^K \sum_{g=1}^G \Bigg(E\Big[Z_{nk}U_{ng}\,\Big|\, \boldsymbol{y},\boldsymbol{\theta}^{(t)}\Big] \log(\phi_k)
E\Big[Z_{nk}(1-U_{ng})\,\Big|\, \boldsymbol{y},\boldsymbol{\theta}^{(t)}\Big]\log(1-\phi_k) \Bigg) \nonumber \\
&+ \sum_{n=1}^N \sum_{k=1}^K \sum_{g=1}^G E\Big[Z_{nk}(1-U_{ng})\,\Big|\,\boldsymbol{y},\boldsymbol{\theta}^{(t)}\Big] \log\Bigg(\frac{\Gamma(y_{ng}+\frac{1}{\alpha_k})}{\Gamma(y_{ng}+1)\Gamma(\frac{1}{\alpha_k})}\left(\frac{1}{1+\alpha_k\mu_{gk}}\right)^{(\frac{1}{\alpha_k})}\left(1-\frac{1}{1+\alpha_k\mu_{gk}}\right)^{y_{ng}}
\Bigg)
\end{align}

Similarly to Section \ref{ZIP_NOX}, the expected values in (\ref{ZINB_M1}) can be computed via $\hat{Z}_{nk}^{(t)}$ and $\hat{U}_{ng}^{(t)}$ given by: 
\begin{align} \label{Z_nk1_ZINB}
\hat{Z}_{nk}^{(t)}&=E\Big[Z_{nk}\,|\,\boldsymbol{y}\,,\,\boldsymbol{\theta}^{(t)}\Big]
=\frac{\pi_k^{(t)}\prod_{g=1}^G p\Big(y_{ng}\,|\,\alpha_k^{(t)},\mu_{gk}^{(t)},\phi_k^{(t)}\Big)}{\sum_{j=1}^K \pi_{j}^{(t)}\prod_{g=1}^G p\Big(y_{ng}\,|\,\alpha_j^{(t)},\mu_{gj}^{(t)},\phi_j^{(t)}\Big)}\mbox{,}
\intertext{and}
\label{U_ng1_ZINB}
\hat{U}_{ngk}^{(t)}
&= \begin{dcases*}
\frac{\phi_k^{(t)}}{ \Bigg(\phi_k^{(t)}+(1-\phi_k^{(t)})\Big(\frac{1}{1+\alpha_k^{(t)}\mu_{gk}^{(t)}}\Big)^{\frac{1}{\alpha_k^{(t)}}}\Bigg)},
& if $y_{ng}=0$, \\
0, & if $y_{ng}=1,2,\dotsc\:$.
\end{dcases*}
\end{align}
Thus, using $\hat{Z}_{nk}^{(t)}$ and $\hat{U}_{ngk}^{(t)}$ from Equations (\ref{Z_nk1_ZINB}) and (\ref{U_ng1_ZINB}), we can rewrite (\ref{ZINB_M1}) as: 
\begin{align}\label{ZINB_Q1}
Q(\boldsymbol{\theta};\boldsymbol{\theta}^{(t)})&=Q_1(\boldsymbol{\pi};\boldsymbol{\pi}^{(t)})+Q_2(\boldsymbol{\phi};\boldsymbol{\phi}^{(t)})+Q_3((\boldsymbol{\mu},\boldsymbol{\alpha});(\boldsymbol{\mu}^{(t)},\boldsymbol{\alpha}^{(t)})),
\end{align}
where the form of $Q_1(\boldsymbol{\pi}\,;\,\boldsymbol{\pi}^{(t)})$ and $Q_2(\boldsymbol{\phi}\,;\,\boldsymbol{\phi}^{(t)})$ remain as in (\ref{Q1_b}) and (\ref{Q2_b}), and 
\begin{align*}
Q_3\Big((\boldsymbol{\mu}\,,\,\boldsymbol{\alpha});(\boldsymbol{\mu}^{(t)},\boldsymbol{\alpha}^{(t)})\Big)= \sum_{n=1}^N \sum_{k=1}^K \sum_{g=1}^G \Bigg[\hat{Z}_{nk}^{(t)} (1-\hat{U}_{ngk}^{(t)})\Bigg\{y_{ng}\,\log\Big(\frac{\alpha_k\mu_{gk}}{1+\alpha_k\mu_{gk}}\Big)-\frac{1}{\alpha_k}\log(1+\alpha_k\mu_{gk})\\
+\log\Gamma(y_{ng}+\frac{1}{\alpha_k})-\log\Gamma(y_{ng}+1)-\log\Gamma\left(\frac{1}{\alpha_k}\right)
\Bigg\}\Bigg].
\end{align*}
\textbf{M-step:}
In this step, we find the updated parameters $\boldsymbol{\theta}^{(t+1)}=(\boldsymbol{\alpha}^{(t+1)},\boldsymbol{\mu}^{(t+1)},\boldsymbol{\phi}^{(t+1)},\boldsymbol{\pi}^{(t+1)})$ that maximize $Q(\boldsymbol{\theta};\boldsymbol{\theta}^{(t)})$ given in Equation (\ref{ZINB_Q1}). The updated estimates of $\pi_k^{(t+1)}$ and $\phi_k^{(t+1)}$ have the same form as in (\ref{P_1}) and (\ref{Phi_1}), respectively. The updated $\mu_{gk}^{(t+1)}$ is obtained as follows:
\begin{equation}\label{mu_1}
   \mu_{gk}^{(t+1)}= \frac{\sum_{n=1}^N  \hat{Z}_{nk}^{(t)}(1-\hat{U}_{ngk}^{(t)}) y_{ng}}{\sum_{n=1}^N \hat{Z}_{nk}^{(t)}(1-\hat{U}_{ngk}^{(t)})}.
\end{equation}

\indent Finally, to reach the updated $\alpha_k^{(t+1)}$, we consider the expectation-conditional maximization (ECM) algorithm by fixing $\mu_{gk}$ at $\mu_{gk}^{(t+1)}$ and obtaining $\alpha_k^{(t+1)}$ as the solution of the following equation:
\begin{align} \label{alpha_1}
& \frac{\partial\,Q_3\Big((\boldsymbol{\mu},\boldsymbol{\alpha})  ;\,(\boldsymbol{\mu^{(t)}},\boldsymbol{\alpha}^{(t)})\Big)}{\partial\,\alpha_k} \nonumber \\
&= \sum_{n=1}^N\sum_{g=1}^G\hat{Z}_{nk}^{(t)}(1-\hat{U}_{ngk}^{(t)})\left[\frac{1}{\alpha_k^2}\left(\log\left(1+\alpha_k\mu_{gk}^{(t+1)}\right)
+ \frac{\alpha_k(y_{ng}-\mu_{gk}^{(t+1)})}{(1+\alpha_k\mu_{gk}^{(t+1)})}+\psi\left(y_{ng}+\frac{1}{\alpha_k}\right)-\psi\left(\frac{1}{\alpha_k}\right)\right)\right]
= 0,
\end{align}
where $\psi(\cdot)$ is the digamma function, which is defined as the derivative of the natural logarithm of the gamma function $\Gamma(\cdot)$, that is, $\psi(x) = \displaystyle\frac{\mathrm{d}}{\mathrm{d}x}\log \Gamma (x)$ \citep{Negative_binomial}. However, there is no closed-form solution for (\ref{alpha_1}); therefore, a numerical optimization algorithm, such as Newton--Raphson or Fisher scoring, must be applied.

Algorithm \ref{alg::ZINB_NoX} in the Appendix shows the ECM algorithm to obtain the updated parameter estimates for the ZINB mixture model without covariates.

\subsubsection{EM for the ZINB mixture model with covariates}\label{ZINB_X}

In the case of a ZINB mixture model with covariates, we assume that the log-link function for the rate parameters is as in \eqref{full_lambda} in Section \ref{ZIP_X}; that is: 
\begin{equation}\label{full_mu}
    \log\mu_{ngk} = \log T_n +\rho_{gk}+\beta_{0g}+\sum_{p=1}^P \beta_{pg} x_{np},
\end{equation}
for $n=1,\dots,N$, $g=1,\dots,G$, $k=1,\dots,K$, and $p=1,\dots,P$, with $T_n$, $\beta_{0g}$, $\rho_{gk}$, $x_{np}$, and $\beta_{pg}$ as in (\ref{full_lambda}).

Let $\boldsymbol{\theta}=(\boldsymbol{\pi},\boldsymbol{\phi}, \boldsymbol{\alpha},\boldsymbol{\rho},\boldsymbol{\beta}_0,\boldsymbol{\beta})$,
where $\boldsymbol{\pi}=(\pi_1,\dots,\pi_k)^T$, $\boldsymbol{\phi}=(\phi_1,\dots,\phi_k)^T$, $\boldsymbol{\alpha}=(\alpha_1,\dots,\alpha_k)^T$, $\boldsymbol{\beta}_0=(\beta_{01},\dots,\beta_{0G})^T$, $\boldsymbol{\beta}=(\beta_{11},\dots,\beta_{P1},\dots,\beta_{1G},\dots,\beta_{PG})^T$, and $\boldsymbol{\rho}=(\rho_{11},\dots,\rho_{G1},\dots,\rho_{1K},\dots,\rho_{GK})^T$.
In this case, the complete-data log-likelihood can be written as:
\begin{align*}
\ell(\boldsymbol{\theta}\,|\,\boldsymbol{y},\boldsymbol{x},\boldsymbol{z},\boldsymbol{u})
= \sum_{n=1}^N \sum_{g=1}^G \sum_{k=1}^K \Big[ z_{nk}  \log\pi_k +z_{nk} u_{ng}\log\phi_k +z_{nk}(1-u_{ng})\,\log(1-\phi_k) + \\
+ z_{nk}(1-u_{ng}) \,\log(p(y_{ng}\,|\,\rho_{gk},\beta_{0g},\beta_{pg},\alpha_k)) \Big].
\end{align*}

\noindent \textbf{E-Step:} The conditional expectation of the complete-data log-likelihood given the observed data and current parameter estimates is as follows:
\begin{align}\label{Q_ZINB_1}
Q(\boldsymbol{\theta};\boldsymbol{\theta}^{(t)})
=& E\Big[\ell(\boldsymbol{\theta}\,|\, \boldsymbol{y},\boldsymbol{x},\boldsymbol{z},\boldsymbol{u})\,\Big|\,\boldsymbol{y},\boldsymbol{x},\boldsymbol{\theta}^{(t)}\Big] \nonumber \\
=& \sum_{n=1}^N \sum_{k=1}^K E\Big[Z_{nk}\,|\,\boldsymbol{y},\boldsymbol{x},\boldsymbol{\theta^{(t)}}\Big] \log(\pi_k) \nonumber \\
&+ \sum_{n=1}^N \sum_{g=1}^G \sum_{k=1}^K \Big(E\Big[Z_{nk}U_{ng}\,|\,\boldsymbol{y},\boldsymbol{x},\boldsymbol{\theta}^{(t)}\Big] \log(\phi_k)+E\Big[Z_{nk}(1-U_{ng})\,|\,\boldsymbol{y},\boldsymbol{x},\boldsymbol{\theta}^{(t)}\Big]\log(1-\phi_k)\Big) \nonumber \\
&+ \sum_{n=1}^N \sum_{g=1}^G \sum_{k=1}^K  E\Big[Z_{nk}(1-U_{ng})\,|\,\boldsymbol{y},\boldsymbol{x},\boldsymbol{\theta}^{(t)}\Big] \nonumber \\
&\times \log\Bigg\{\frac{\Gamma(y_{ng}+\frac{1}{\alpha_k})}{\Gamma(y_{ng}+1)\Gamma(\frac{1}{\alpha_k})}
\times \left(\frac{1}{1+\alpha_k\exp\{\log T_n + \rho_{gk}+\beta_{0g}+\sum_{p=1}^P \beta_{pg} x_{np}\}}\right)^{\frac{1}{\alpha_k}} \nonumber\\
&\times \left(1-\frac{1}{1+\alpha_k \exp\{\log T_n + \rho_{gk}+\beta_{0g}+\sum_{p=1}^P \beta_{pg} x_{np}\}}\right)^{y_{ng}}
\Bigg\}
\end{align}   
As described in previous sections, we can calculate the expectations in (\ref{Q_ZINB_1}) using $\hat{Z}_{nk}$ and $\hat{U}_{ngk}$ given by:
\begin{align}\label{Z_nk2_ZINB}
\hat{Z}_{nk}^{(t)}
&= E\Big[Z_{nk}\,|\,\boldsymbol{y},\boldsymbol{x},\boldsymbol{\theta}^{(t)} \Big]=\frac{\pi_k^{(t)}\prod_{g=1}^G p(y_{ng}\,|\,\phi_k^{(t)},\alpha_k^{(t)},\rho_{gk}^{(t)},\beta_{0g}^{(t)},\beta_{pg}^{(t)})}{\sum_{k=1}^K\pi_k^{(t)}\prod_{g=1}^G p(y_{ng}\,|\,\phi_k^{(t)},\alpha_k^{(t)},\rho_{gk}^{(t)},\beta_{0g}^{(t)},\beta_{pg}^{(t)})},
\intertext{and}
\label{U_ng2_ZINB}
\hat{U}_{ngk}^{(t)}
&= \begin{dcases*}
\frac{\phi_k^{(t)}}{\Bigg(\phi_k^{(t)}+(1-\phi_k^{(t)})\Big(\frac{1}{1+\alpha_k^{(t)}\mu_{ngk}^{(t)}}\Big)^{\frac{1}{\alpha_k^{(t)}}}\Bigg)},
& if $y_{ng}=0$, \\
0, & if $y_{ng}=1,2,\dotsc$,
\end{dcases*}  
\end{align}
where $\mu_{ngk}^{(t)}=\exp\Big\{\log T_n +\rho_{gk}^{(t)}+\beta_{0g}^{(t)}+\sum_{p=1}^P \beta_{pg}^{(t)} x_{np}\Big\}$.

Using $\hat{Z}_{nk}$ and $\hat{U}_{ngk}$ as in (\ref{Z_nk2_ZINB}) and (\ref{U_ng2_ZINB}), respectively, we can rewrite $Q(\boldsymbol{\theta};\boldsymbol{\theta}^{(t)})$ as follows: 
\begin{align}\label{eq:Q_ZINB_cov}
Q(\boldsymbol{\theta}\,;\,\boldsymbol{\theta}^{(t)})&=Q_1(\boldsymbol{\pi}\,;\,\boldsymbol{\pi}^{(t)})+Q_2(\boldsymbol{\phi}\,;\,\boldsymbol{\phi}^{(t)})+Q_3((\boldsymbol{\beta}_{0},\boldsymbol{\rho},\boldsymbol{\beta},\boldsymbol{\alpha})\,;\,(\boldsymbol{\beta}_{0}^{(t)},\boldsymbol{\rho}^{(t)},\boldsymbol{\beta}^{(t)},\boldsymbol{\alpha}^{(t)})),
\end{align}
where again $Q_1(\boldsymbol{\pi}\,;\,\boldsymbol{\pi}^{(t)})$ and $Q_2(\boldsymbol{\phi}\,;\,\boldsymbol{\phi}^{(t)})$ have the form of (\ref{Q1_b}) and (\ref{Q2_b}), respectively, and
\begin{align}\label{FullQ_3_ZINB}
& Q_3\Big((\boldsymbol{\alpha},\boldsymbol{\rho},\boldsymbol{\beta}_0,\boldsymbol{\beta});(\boldsymbol{\alpha}^{(t)},\boldsymbol{\rho}^{(t)},\boldsymbol{\beta}_0^{(t)},\boldsymbol{\beta}^{(t)}
\Big)
= \sum_{n=1}^N \sum_{g=1}^G \sum_{k=1}^K \hat{Z}_{nk}^{(t)} (1-\hat{U}_{ngk}^{(t)}) \nonumber \\
&\times \log\Bigg\{\frac{\Gamma(y_{ng}+\frac{1}{\alpha_k})}{\Gamma(y_{ng}+1)\Gamma(\frac{1}{\alpha_k})}
\times \left(\frac{1}{1+\alpha_k\exp\{\log T_n+\rho_{gk}+\beta_{0g}+\sum_{p=1}^P \beta_{pg} x_{np}\}}\right)^{\frac{1}{\alpha_k}} \nonumber \\
&\times \left(1-\frac{1}{1+\alpha_k \exp\{\log (T_n)+\rho_{gk}+\beta_{0g}+\sum_{p=1}^P \beta_{pg} x_{np}\}}\right)^{y_{ng}}
 \Bigg\}.
\end{align}

\paragraph{M-Step:} In this step, we maximize $Q(\boldsymbol{\theta}\,;\,\boldsymbol{\theta}^{(t)})$ in (\ref{eq:Q_ZINB_cov}) with respect to each parameter in $\boldsymbol{\theta}$ to find the updated parameter estimates. The updated estimates of $\pi_k$ and $\phi_k$ can be obtained in closed form as in (\ref{P_1}) and (\ref{Phi_1}), respectively. However, no closed-form solution exists for $\alpha_k$, $\beta_{0g}$, $\rho_{gk}$, and $\beta_{gp}$; therefore, we apply the ECM algorithm along with the Newton--Raphson optimization method to find $\alpha_k^{(t+1)}$, $\beta_{0g}^{(t+1)}$, $\rho_{gk}^{(t+1)}$, and $\beta_{gp}^{(t+1)}$. 

We summarize the ECM algorithm to obtain the updated parameter estimates for the ZINB mixture model with covariates in Algorithm \ref{alg::ZINB_X} in the Appendix. As in the Poisson case, a simpler model than in (\ref{full_mu}) can be considered when there is a size factor but no covariates, that is, we can model $\mu_{ngk}$ as
\begin{equation}\label{reduced_mu}
    \log \mu_{ngk} = \log T_n +\rho_{gk}+\beta_{0g}.
\end{equation}
In this case, in the E-step, $Q_1$ and $Q_2$ have the same form as in (\ref{Q1_b}) and (\ref{Q2_b}), but $Q_3$ now becomes:
\begin{align}\label{reducedQ_3_ZINB}
Q_3\Big((\boldsymbol{\alpha},\boldsymbol{\beta}_{0},\boldsymbol{\rho})\,;\,(\boldsymbol{\alpha}^{(t)},\boldsymbol{\beta}_{0}^{(t)},\boldsymbol{\rho}^{(t)}\Big)
&= \sum_{n=1}^N \sum_{g=1}^G \sum_{k=1}^K \hat{Z}_{nk}^{(t)} (1-\hat{U}_{ngk}^{(t)}) \nonumber \\
&\times \log\Bigg\{\frac{\Gamma(y_{ng}+\frac{1}{\alpha_k})}{\Gamma(y_{ng}+1)\Gamma(\frac{1}{\alpha_k})} \times \left(\frac{1}{1+\alpha_k \times \exp\{\log (T_n)+\rho_{gk}+\beta_{0g}\}}\right)^{\frac{1}{\alpha_k}} \nonumber \\
&\times \left(1-\frac{1}{1+\alpha_k \times \exp\{\log (T_n)+\rho_{gk}+\beta_{0g}\}}\right)^{y_{ng}}
\Bigg\}.
\end{align}

The updated estimates of $\pi_k$ and $\phi_k$ are as in (\ref{P_1}) and (\ref{Phi_1}), respectively, and  the updated estimates of $\beta_{0g}$, $\rho_{gk}$, and $\alpha_k$ (or $\nu_k$) can also be obtained via Newton--Raphson within the ECM algorithm. 

Similarly to the ZIP case, in the code implementation, to also avoid identifiability issues when estimating the parameters of the ZINB mixture model with covariates, one can consider $\beta_{gk}=\beta_{0g}+\rho_{gk}$ with the restriction that $\sum_{k=1}^K \rho_{gk}=0$. For more details, see the end of Section \ref{ZIP_X}.

\section{Simulation studies}\label{Results}
In this Section, for each model introduced in Section \ref{sec:method}, we conduct simulation studies to assess the performance of our proposed EM algorithm under various scenarios by varying different parameters and hyperparameters. For the hyperparameters, we vary the number of subjects ($N$) (the number of rows in the $\mathbf{y}$ matrix in Eq. (\ref{Y_matrix_CH4})), the number of observations ($G$) (the number of columns in the $\mathbf{y}$ matrix), and the number of clusters ($K$). We note that the number of clusters is fixed in the EM algorithm. However, the optimal $K$ can be found using a criterion such as the Akaike Information Criterion (AIC), the Bayesian Information Criterion (BIC), or the Integrated Complete Likelihood (ICL) \citep{Mclachlan2007}. In Section \ref{ch:data_analysis}, when analyzing publicly available datasets, we found the best $K$ using AIC. For the parameters, we consider different values of $\pi_1, \ldots, \pi_K$, the cluster assignment probabilities, $\phi_1,\ldots,\phi_K$, the probabilities of always zero, and $\nu_1,\ldots,\nu_K$, the size parameters for the case of a mixture of ZINB distributions. We also examine different values for the rate parameters for the cases with and without covariates for the mixture of ZIP and ZINB distributions. Section 1.1 of the Supplementary Material presents some of the computer and software specifications for our simulation studies. 

\subsection{Performance metrics}

In what follows, the different metrics and plots used to assess the performance of the proposed EM algorithm are introduced. For evaluating the performance regarding the parameters $\pi_1,\dots,\pi_k$ and $\phi_1,\dots,\phi_k$, the means and standard deviations of the obtained EM estimates along with boxplots are computed across the different simulation scenarios. To evaluate the performance regarding the estimation of the rate parameters $\lambda_{gk}$'s or $\mu_{gk}$'s (case without covariates) and $\beta_{0g}$'s, $\beta_{pg}$'s, and $\rho_{gk}$'s (case with covariates), the mean squared error (MSE) or the median absolute deviation (MAD) are applied. See Section 1.2 of the Supplementary Material for details on how we calculate the MSE and MAD in our simulations. For the size parameters in the ZINB case, we present boxplots, means, and standard deviations of the EM estimates. The V-measure \citep{VMeasure} is used to evaluate the clustering performance, i.e., how well the clustering performs compared to the true assigned clusters of each data set. See also Section 1.2 of the Supplementary Material for more details on performance metrics as well as the V-measure.

\subsection{Simulation results for the ZIP mixture model without covariates} \label{sec:simZIPNoX}

We study six different simulation scenarios considering the ZIP mixture model without covariates presented in Section \ref{ZIP_NOX}. In each scenario, we generate data based on the ZIP model without covariates, varying one of the model parameters or hyperparameters while holding the others fixed. We then fit the data considering a ZIP model without covariates and the proposed EM algorithm. In Scenarios $1$ and $2$, we vary the number of subjects ($N$) and the number of observations ($G$), respectively. In Scenario $3$, we vary the number of clusters ($K$). In Scenario $4$, we investigate the effect of changing the cluster assignment probabilities ($\pi_k$'s) from balanced to unbalanced cases. In Scenario $5$, we vary the similarities among clusters by changing the $\lambda_{gk}$'s. Finally, the effect of changes in the probabilities of always zero, $\phi_k$'s, is studied in Scenario $6$. 

Table \ref{t:setting} summarizes each of the proposed simulation scenarios for the case of a ZIP mixture model without covariates, and it shows which parameters and hyperparameters vary in each scenario (indicated by the $\star$ symbol) along with the ones that are kept fixed.
We simulate $S=256$ datasets for each scenario and apply the proposed EM algorithm (Algorithm \ref{alg::ZIP_No_X} in Section \ref{ZIP_NOX}) to each of these datasets, setting the initial parameter values to the true parameter values to speed up the computation process.

In Sections \ref{sec:sim_zipNOX_scenario1} and \ref{sec:sim_zipNOX_scenario2}, we present the simulation results for Scenarios 1 and 2, respectively. Results from Scenarios 3 to 6 are available from Section 1.3.3 to Section 1.3.6 of the Supplementary Material.

\begin{table}
\centering
\caption{Settings used for each simulation study scenario. The $\star$ indicates the parameter or hyperparameter that varies in each scenario.}
\label{t:setting}
\begin{tabular}{c*{5}{r}}
 \toprule
Scenario    &     $N$ &     $G$ &     $K$ & $\phi_k$ & $\pi_k$ \\
\midrule
1           & $\star$ &     120 &       3 &      0.1 & $\rfrac{1}{K}$ \\ 
2           &    1200 & $\star$ &       3 &      0.1 & $\rfrac{1}{K}$ \\ 
3           &    1200 &     120 & $\star$ &      0.1 & $\rfrac{1}{K}$ \\ 
4           &    1200 &     120 &       2 &      0.1 & $\star$ \\ 
5$^{\ddag}$ &    1200 &     120 &       2 &      0.1 & $\rfrac{1}{K}$ \\
6           &    1200 &     120 &       3 &  $\star$ & $\rfrac{1}{K}$ \\
   \bottomrule
\end{tabular}
\\{\footnotesize $^{\ddag}$ The settings are kept the same but we vary
the similarities among clusters by changing the $\lambda_{gk}$'s.}
\end{table}

\subsubsection{Scenario 1 - ZIP mixture model without covariates}\label{sec:sim_zipNOX_scenario1}
In this scenario, we vary $N$ according to six different values (cases), while all other parameters are kept fixed as shown in Table \ref{t:sim11} below.
Three distinct values are chosen for the rate parameters ($\lambda_{kg}$'s) and we repeat the same value for a third of the number of observations in each cluster (i.e., $\frac{G}{3}=40$ times), in a way that the rate parameters are distinct for each observation and across clusters. For this scenario, we choose $\lambda_1=5$, $\lambda_2=10$, and $\lambda_3=15$ and we use the following matrix for generating the simulated data sets:
\begin{equation*}
\mathbf{\lambda} = \begin{pmatrix}
\lambda_1 & \cdots & \lambda_1 & \lambda_2 & \cdots & \lambda_2 & \lambda_3 & \cdots & \lambda_3\\
\lambda_2 & \cdots & \lambda_2 & \lambda_3 & \cdots & \lambda_3 & \lambda_1 & \cdots & \lambda_1\\
\lambda_3 & \cdots & \lambda_3 & \lambda_1 & \cdots & \lambda_1 & \lambda_2 & \cdots & \lambda_2
\end{pmatrix}.
\end{equation*}
\textcolor{black}{Figure 1 
of the Supplementary Material} shows an example of simulated data for 
Case 3 in Table \ref{t:sim11}.


\begin{table}
	\centering
	\caption{\textbf{ZIP mixture model without covariates.} \scenario{1} Values chosen for the number of subjects $N$ in each of five cases along with the fixed parameters used to simulate the datasets under a ZIP mixture model without covariates.}
	\label{t:sim11}
	\begin{tabular}{c*{5}{r}}
		\toprule
		Case & $N$ & $G$ & $K$ & $\phi_k$ & $\pi_k$ \\
		\midrule
		1 &   12 \\
		2 &   60 \\
		3 &  120 & 120 & 3 & 0.1 & \(\rfrac{1}{K}\) \\
		4 &  600 \\
		5 & 1200 \\
		\bottomrule
	\end{tabular}
\end{table}

\begin{figure}
\centering
\includegraphics{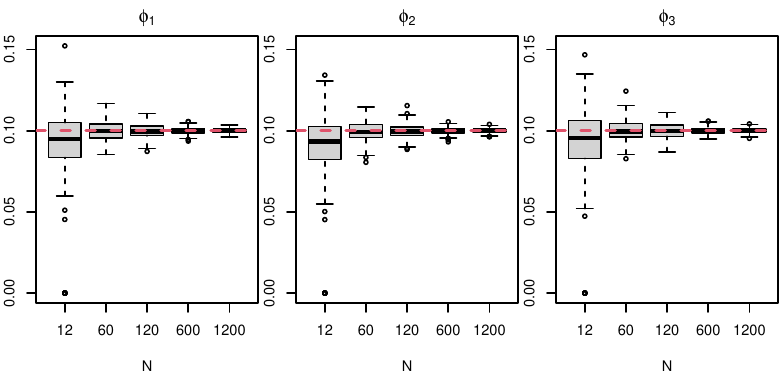}\\[0.2cm]
\includegraphics{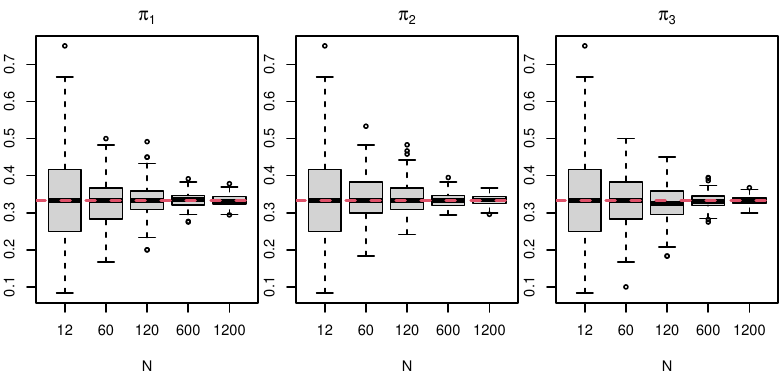}
\caption{\textbf{ZIP mixture model without covariates.} \scenario{1} Boxplots for the estimates of \(\phi_k\) (\textit{top row}) and $\pi_k$ (\textit{bottom row}), for $k=1,\ldots,3$ obtained using the proposed EM algorithm across the datasets simulated from the settings described in Table~\ref{t:sim11}. Red lines correspond to true values.
See also \textcolor{black}{Supplementary Tables 1 and 2 
.}}
\label{f:sim11-phi-pi}
\end{figure}

For this scenario, Figure \ref{f:sim11-phi-pi} and Supplementary Tables 1 and 2 
show that the EM estimates for $\phi_k$ and $\pi_k$, for $k=1,2,$ and 3, are centered around the true values across all the different choices of $N$, except for $\phi_1,\phi_2$ and $\phi_3$ when $N=12$.
Furthermore, as $N$ increases, the standard deviations of these estimates decrease as desired. Table \ref{t:sim11-lambda} demonstrates that the MSE for estimating the $\lambda_{gk}$'s decreases while $N$ increases. Moreover, according to \textcolor{black}{Supplementary Figure 2 
}, the clustering performance measured by the V-measures is deemed optimal (V-measure $= 1$) except for the lowest value of $N=12$, which results in some misclassifications. Finally, we can see from \textcolor{black}{Supplementary Figures 3 and 4  
and Supplementary Tables 3 and 4 
} 
that although the computation time increases, as $N$ increases, the number of iterations until convergence decreases for the first three cases and then stabilizes for the larger $N$ cases.

\begin{table}
\centering
\caption{\textbf{ZIP mixture model without covariates.} \scenario{1} Mean squared error across observations and simulated datasets for the EM estimates of the $\lambda_{gk}$'s for each cluster $k$ and each $N$ according to the settings described in Table~\ref{t:sim11}.} 
\label{t:sim11-lambda}
\begin{tabular}{lrrrrr}
  \toprule
  & \multicolumn{5}{c}{$N$}\\
 \cmidrule(l){2-6}
 $k$ & 12 & 60 & 120 & 600 & 1200\\
 \midrule
1 & 4.66910 & 0.58775 & 0.28343 & 0.05595 & 0.02819 \\ 
  2 & 4.18363 & 0.57776 & 0.28229 & 0.05559 & 0.02740 \\ 
  3 & 4.48772 & 0.57781 & 0.29115 & 0.05579 & 0.02800 \\ 
   \bottomrule
\end{tabular}
\end{table}

\subsubsection{Scenario 2 -- ZIP mixture model without covariates}\label{sec:sim_zipNOX_scenario2}
In this scenario, the number of observations $G$ varies while the other parameters are kept fixed, as shown in Table \ref{t:sim12}. The number of clusters, the cluster assignment probabilities, the probabilities of always zero, and the $\lambda_{gk}$'s are fixed to values similar to those in Scenario 1. We fix $N=1200$ for this scenario, as seen in Table \ref{t:sim12}. 

\begin{table}
	\centering
	\caption{\textbf{ZIP mixture model without covariates.} \scenario{2} Values chosen for the number of observations $G$ in each of five cases along with the fixed parameters used to simulate the datasets.}
	\label{t:sim12}
	\begin{tabular}{c*{5}{r}}
		\toprule
		Case & $N$ & $G$ & $K$ & $\phi_k$ & $\pi_k$ \\
		\midrule
		1 &      &   12 \\
		2 &      &   60 \\
		3 & 1200 &  120 & 3 & 0.1 & \(\rfrac{1}{K}\) \\
		4 &      &  600 \\
		5 &      & 1500 \\
		\bottomrule
	\end{tabular}
\end{table}

\begin{figure}
\centering
\includegraphics{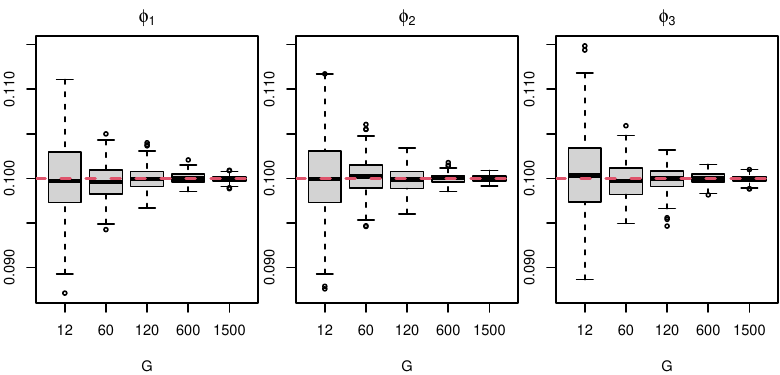}\\[0.2cm]
\includegraphics{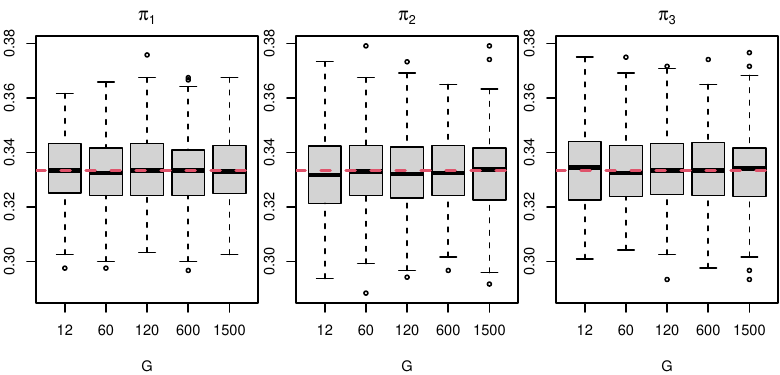}
\caption{\textbf{ZIP mixture model without covariates.} \scenario{2} Boxplots for the estimates of \(\phi_k\) (\textit{top row}) and $\pi_k$ (\textit{bottom row}), for $k=1,\ldots,3$ obtained using the proposed EM algorithm across the datasets simulated from the settings described in Table~\ref{t:sim12}. Red lines correspond to true values. See also \textcolor{black}{Supplementary Tables
5 and 6.}}
\label{f:sim12-phi-pi}
\end{figure}

Figure \ref{f:sim12-phi-pi} and \textcolor{black}{Supplementary Table 5
} show that as $G$ increases, the standard deviation of the estimates of each $\phi_k$ decreases, while the estimates remain centred around the true value of $0.1$. For the estimates of $\pi_k$, according to Figure \ref{f:sim12-phi-pi} and \textcolor{black}{Supplementary Table 6
}, their standard deviations remain somewhat the same when varying $G$ across clusters. In addition, estimates of $\pi_k$ are unbiased, as in Scenario 1. Table \ref{t:sim12-lambda} shows that the MSE of the estimates of the $\lambda_{gk}$'s for each cluster remains almost the same while varying $G$. Furthermore, the resulting V-measure values (\textcolor{black}{Supplementary Figure 5 
}) are equal to one for $G=60,120,600,1500$. However, for $G=12$, a few misclassifications lead to V-measure values slightly less than one. The number of iterations remains constant (see \textcolor{black}{Table 7 
and Figure 6
in the Supplementary Material}), and the total computing time increases as $G$ increases (see \textcolor{black}{Table 8 
and Figure 7
in the Supplementary Material}).  



\begin{table}[ht]
\centering
\caption{\textbf{ZIP mixture model without covariates.} \scenario{2} Mean squared error across observations and simulated datasets for the EM estimates of the $\lambda_{gk}$'s for each cluster $k$ and each $G$ according to the settings described in Table~\ref{t:sim12}.} 
\label{t:sim12-lambda}
\begin{tabular}{lrrrrr}
  \toprule
  & \multicolumn{5}{c}{$G$}\\
 \cmidrule(l){2-6}
 $k$ & 12 & 60 & 120 & 600 & 1500\\
 \midrule
1 & 0.02826 & 0.02804 & 0.02796 & 0.02823 & 0.02792 \\ 
  2 & 0.02823 & 0.02838 & 0.02794 & 0.02799 & 0.02810 \\ 
  3 & 0.02804 & 0.02879 & 0.02777 & 0.02791 & 0.02792 \\ 
   \bottomrule
\end{tabular}
\end{table}

\subsection{Simulation results for the ZIP mixture model with \textcolor{black}{an offset or a size factor}}\label{sec:sim:ZIP_size_factor}

In the simulation scenarios of this section, we model the rate parameter (expected read count) via a log link function as  $\log(\lambda_{ngk})=\log(T_n)+\beta_{0g}+\rho_{gk}$, which was introduced earlier in Section \ref{ZIP_X}. So, to simulate data under different scenarios, we construct the rate parameter by simulating \textcolor{black}{size factors}, $T_n$, for $n=1\ldots, N$, from a normal distribution with parameters $\mu=1000$, $\sigma=100$, and we fix $\beta_{0g}$ at a value of one for all observations. For most of the cases, we choose three distinct values ($-0.6$, 0 and 0.6) for the cluster effects ($\rho_{gk}$'s), and we repeat the same value for a third of the number of observations in each cluster (i.e., $\frac{G}{3}$ times), in a way that the rate parameters for each observation are distinct across clusters. Note that, to avoid identifiability issues,  we consider the restriction of $\sum_{k=1}^K \rho_{gk}=0$ to select the values of $\rho_{gk}$. 

In this case (ZIP mixture model with a size factor), we study six different scenarios, each with 256 simulated datasets. Scenarios 1, 2, and 3 involve varying the values of $N$, $G$, and $K$, respectively. In Scenario 4, we consider two cases: Case 1 has initial parameter values matching the true values used to generate the data, while Case 2 has initial parameter values obtained through $K$-means clustering. Scenario 5 focuses on different probabilities of cluster assignments, while Scenario 6 varies the probabilities of always zero and compares the results. Note that, except for Case 2 in Scenario 4, we set the initial parameter values to the true values to speed up computation, as we did in Section \ref{sec:simZIPNoX}. In what follows, we present the simulation results for Scenario 1. Results from Scenarios 2 to 6 are available in \textcolor{black}{Sections 1.4.2 through 1.4.6 of the Supplementary Material}. 

\begin{figure}
\centering
\includegraphics{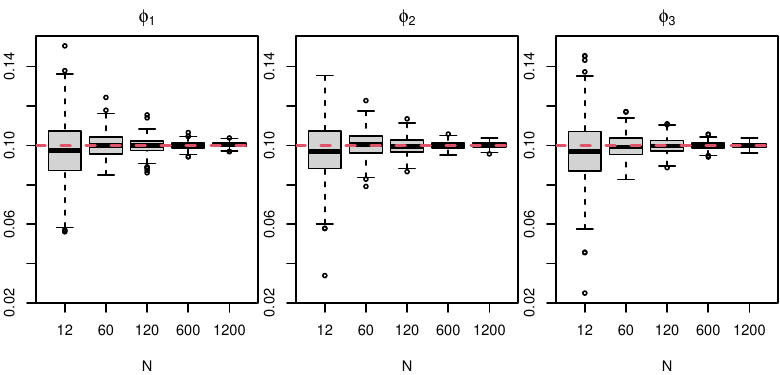}\\[0.2cm]
\includegraphics{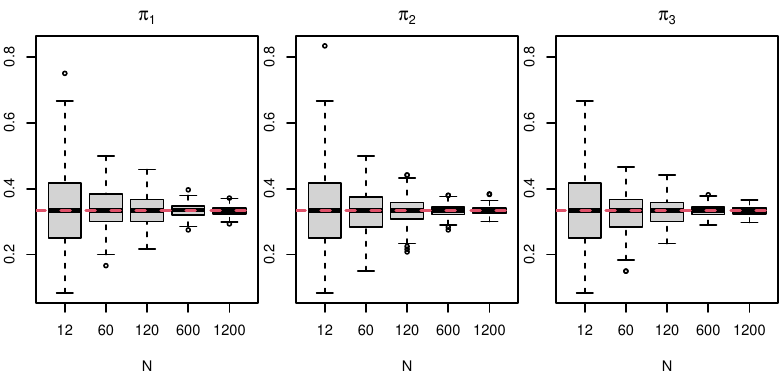}
\caption{\textbf{ZIP mixture model with a size factor.} \scenario{1} Boxplots for the estimates of \(\phi_k\) (\textit{top row}) and $\pi_k$ (\textit{bottom row}), for $k=1,\ldots,3$ obtained using the proposed EM algorithm across the datasets simulated from the settings described in Table~\ref{t:sim21}. Red lines correspond to true values.
See also \textcolor{black}{Supplementary Tables 33 and 34
}.}
\label{f:sim21-phi-pi}
\end{figure}

 In Scenario 1, similar to the case without covariates in Section \ref{sec:sim_zipNOX_scenario1}, we vary $N$ while all other parameters and hyperparameters are kept fixed. See Table \ref{t:sim21} for the parameter setting used to generate data under this scenario. Figure \ref{f:sim21-phi-pi} and \textcolor{black}{Supplementary Tables 33 and 34
} show that estimated probabilities of always zero ($\hat{\phi}_k$) and estimated cluster assignment's probabilities ($\hat{\pi}_k$) are  approximately around their true values and as $N$ increases, their standard deviations decrease. For the estimated $\rho_{gk}$ and $\beta_{0g}$, we  consider the median absolute deviation (MAD) and we can observe from Tables \ref{t:sim21-rho} and \ref{t:sim21-beta0} that as $N$ increases, the MADs for both $\rho_{gk}$ and $\beta_{0g}$ decrease. For most cases, the V-measures are one (see \textcolor{black}{Supplementary Figure 22}), except for the case with $N=12$ where there is some misclassification. As expected, the computing times increases while $N$ increases (see \textcolor{black}{Table 35 
and Figure 23 
in the Supplementary Material}). Finally, the required number of iterations to achieve convergence decreases as $N$ grows (see \textcolor{black}{Table 36 
and Figure 24 
in the Supplementary Material}).

\begin{table}
	\centering
	\caption{\textbf{\textcolor{black}{ZIP mixture model with a size factor}.} \scenario{1} Values chosen for the number of subjects $N$ in each case along with the fixed parameters used to simulate the datasets.}
	\label{t:sim21}
	\begin{tabular}{c*{5}{r}}
		\toprule
		Case & $N$ & $G$ & $K$ & $\phi_k$ & $\pi_k$ \\
		\midrule
		1 &   12 \\
		2 &   60 \\
		3 &  120 & 120 & 3 & 0.1 & \(\rfrac{1}{K}\) \\
		4 &  600 \\
		5 & 1200 \\
		\bottomrule
	\end{tabular}
\end{table}

\begin{table}
\centering
\caption{\textbf{ZIP mixture model with a size factor.} \scenario{1} Median absolute deviation for the estimates of $\rho_{gk}$ for each $k$ and each $N$, using the EM algorithm across the datasets simulated from the settings described in Table~\ref{t:sim21}.} 
\label{t:sim21-rho}
\begin{tabular}{lrrrrr}
  \toprule
  & \multicolumn{5}{c}{$N$}\\
 \cmidrule(l){2-6}
 $k$ & 12 & 60 & 120 & 600 & 1200\\
 \midrule
1 & 0.112408550 & 0.045393355 & 0.031839327 & 0.014172495 & 0.009948854 \\ 
  2 & 0.112812275 & 0.044446965 & 0.031538945 & 0.014278217 & 0.010051250 \\ 
  3 & 0.115637525 & 0.045467278 & 0.031932817 & 0.014086308 & 0.010112260 \\ 
   \bottomrule
\end{tabular}
\end{table}

\begin{table}
\centering
\caption{\textbf{ZIP mixture model with a size factor.} \scenario{1} Median absolute deviation for the estimates of $\beta_{0g}$ for each $N$, using the EM algorithm across the datasets simulated from the settings described in Table~\ref{t:sim21}.} 
\label{t:sim21-beta0}
\begin{tabular}{l*{5}{r}}
  \toprule
$N$ & 12 & 60 & 120 & 600 & 1200\\
 \midrule
$\beta_{0g}$ & 0.08286 & 0.03281 & 0.02255 & 0.01017 & 0.00729 \\ 
   \bottomrule
\end{tabular}
\end{table}



\subsection{Simulation scenarios for the ZINB mixture  model without covariates}

In this section, we simulate data from the zero-inflated negative binomial mixture model without covariates (Section \ref{ZINB_NOX}). We consider two scenarios where the number of subjects ($N$) varies in Scenario 1, and in Scenario 2, the number of observations ($G$) varies while holding all other parameters and hyperparameters fixed. For both scenarios, we simulate data from $K=2$ clusters with equal cluster assignment probabilities ($\pi_1=\pi_2=0.5$). The probabilities of always zero are equal to $0.1$ for both clusters ($\phi_1=\phi_2=0.1$). The size parameters for the negative binomial components are $\nu_{1}=5$ and $\nu_{2}=20$. For the negative binomial rate parameters, we considered $\mu_{g1}=\mu_1=5$ and $\mu_{g2}=\mu_2=10$ for all $g$. In what follows, we present the results for Scenario $1$. Results for scenario $2$ are described in Section 1.6.2 of the Supplementary Material. 

\begin{figure}
\centering
\includegraphics[width=0.65\textwidth]{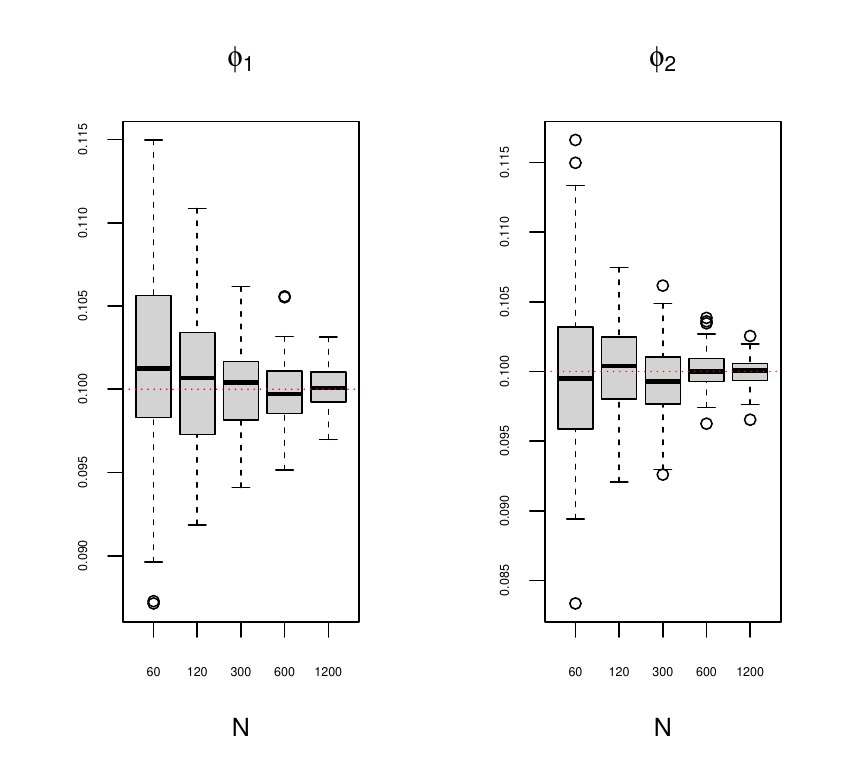}
\includegraphics[width=0.65\textwidth]{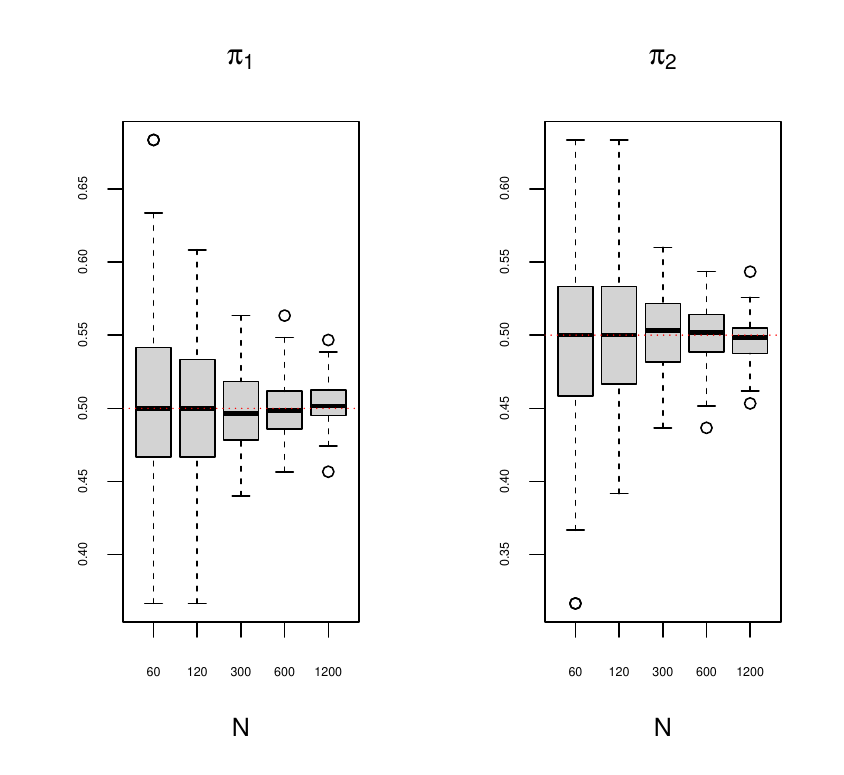}
\caption{\textbf{ZINB mixture  model without covariates.} \scenario{1} Boxplots for the estimates of $\phi_k$ (\textit{top row}) and \(\pi_k\) (\textit{bottom row}) using the EM algorithm across the datasets simulated from the settings described in Table~\ref{t:SC1_ZINB1}. Red lines correspond to true values.
See also \textcolor{black}{Supplementary Tables 81 and 82.}}
\label{f:SC1_ZINB1-phi-pi}
\end{figure}

\begin{figure}
\centering
\includegraphics[width=0.65\textwidth]{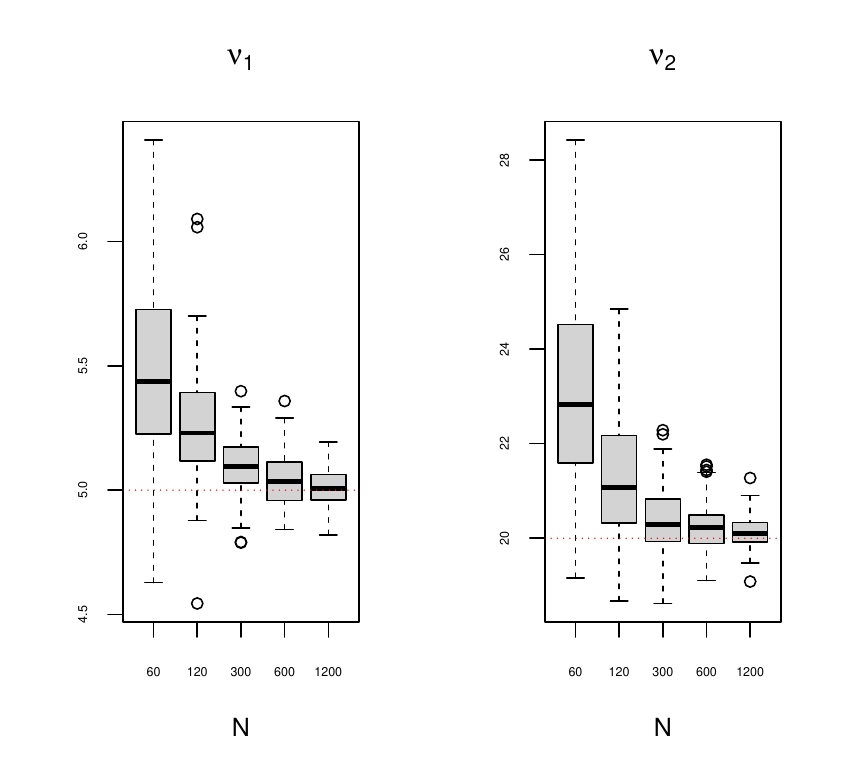}
\caption{\textbf{ZINB mixture  model without covariates.} \scenario{1} Boxplots for the estimates of \(\nu_k\) using the EM algorithm across the datasets simulated from the settings described in Table~\ref{t:SC1_ZINB1}. Red lines correspond to true values.
See also \textcolor{black}{Supplementary Table 83}.}
\label{f:SC1_ZINB1-nu}
\end{figure}

In Scenario 1, the number of subjects ($N$) varies while we fix $G=120$ and all other parameters and hyperparameters according to the setting in Table \ref{t:SC1_ZINB1}. We can observe from Figure \ref{f:SC1_ZINB1-phi-pi} and \textcolor{black}{Supplementary Tables 84 and 85
} that the estimates of $\pi_k$ and $\phi_k$ are close to their true values, and as $N$ increases, the standard deviations decrease. The MSEs for the estimates of the rate parameters decrease as $N$ increases (see Table \ref{t:SC1_ZINB1-mu}).
Furthermore, as shown in Figure \ref{f:SC1_ZINB1-nu} and \textcolor{black}{Supplementary Table 86
}, for both clusters, the bias and the variance in the estimation of the size parameter decrease as $N$ increases. Finally, computing time increases when $N$ increases (see \textcolor{black}{Table 87
and Figure 47
in the Supplementary Material}), and V-measures for all data sets are equal to one.   

\begin{table}
	\centering
	\caption{\textbf{\textcolor{black}{ZINB mixture  model without covariates}.} \scenario{1} Values chosen for the number of subjects $N$ in each case along with the fixed parameters used to simulate the datasets.}
	\label{t:SC1_ZINB1}
	\begin{tabular}{c*{5}{r}}
		\toprule
		Case & $N$ & $G$ & $K$ & $\phi_k$ & $\pi_k$ \\
		\midrule
		1 &   60 \\
            2 & 120\\
		3 &  300 & 120 & 2 & 0.1 & \(\rfrac{1}{K}\) \\
		4 &  600 \\
		5& 1200 \\
		\bottomrule
	\end{tabular}
\end{table}

\begin{table}
\centering
\caption{\textbf{ZINB mixture  model without covariates.} \scenario{1} Mean squared error for the estimates of $\mu_{gk}$ for each $k$ and each $N$, using the EM algorithm across the datasets simulated from the settings described in Table~\ref{t:SC1_ZINB1}.}
\label{t:SC1_ZINB1-mu}
\begin{tabular}{rrrrrr}
  \toprule
  & \multicolumn{5}{c}{$N$}\\
 \cmidrule(l){2-6}
   
$k$ & $60$ & $120$ & $300$ & $600$ & $1200$ \\ 
  \midrule
 1& 0.40832 & 0.19947 & 0.08136 & 0.03927 & 0.01948 \\ 
 2 & 0.57480 & 0.27671 & 0.11284 & 0.05625 & 0.02844 \\ 
   \bottomrule
\end{tabular}
\end{table}

\subsection{Simulation scenario for the ZINB mixture model with \textcolor{black}{an offset or a size factor}}
In this section, we simulate only one scenario for the case of ZINB mixture model with \textcolor{black}{a size factor} when the number of subjects varies as $N=60,120,300,600,$ and $1200$. We choose $G=120$ as the number of observations and $K=2$ as the number of clusters. We simulate $S=100$ datasets, and the true values of the parameters and hyperparameters are set as follows: $\pi=(0.5,0.5)$; $\phi=(0.1,0.2)$; $T_n\mbox{'s}$ are generated from a normal distribution with $\mu=10$ and $\sigma=0.5$; $\beta_{0g}=0.85$ for all $g$; and $\rho_{1g}=(2,\dots,2,-2,\dots,-2)$, and $\rho_{2g}=(-2,\dots,-2,2,\dots,2)$.
Note that, for the cluster effect parameters ($\rho_{gk}$'s), over each cluster, the first half of the observations are assigned one value (either $2$ or $-2$), and the remaining half is assigned another value (either $2$ or $-2$) in such a way that their sums are equal to zero. The setting for this simulation scenario is also shown in Table \ref{t:SC1_ZINB2}.

\begin{table}
	\centering
	\caption{\textbf{\textcolor{black}{ZINB mixture model with a size factor}.} \scenario{1} Values chosen for the number of subjects $N$ in each case along with the fixed parameters used to simulate the datasets.}
	\label{t:SC1_ZINB2}
	\begin{tabular}{c*{5}{r}}
		\toprule
		Case & $N$ & $G$ & $K$ & $\phi_k$ & $\pi_k$ \\
		\midrule
		1 &   60 \\
            2 & 120\\
		3 &  300 & 120 & 2 & ($\phi_1=0.1$,$\phi_2=0.2$) & \(\rfrac{1}{K}\) \\
		4 &  600 \\
		5& 1200 \\
		\bottomrule
	\end{tabular}
\end{table}

\begin{figure}
\centering
\includegraphics[width=0.65\textwidth]
{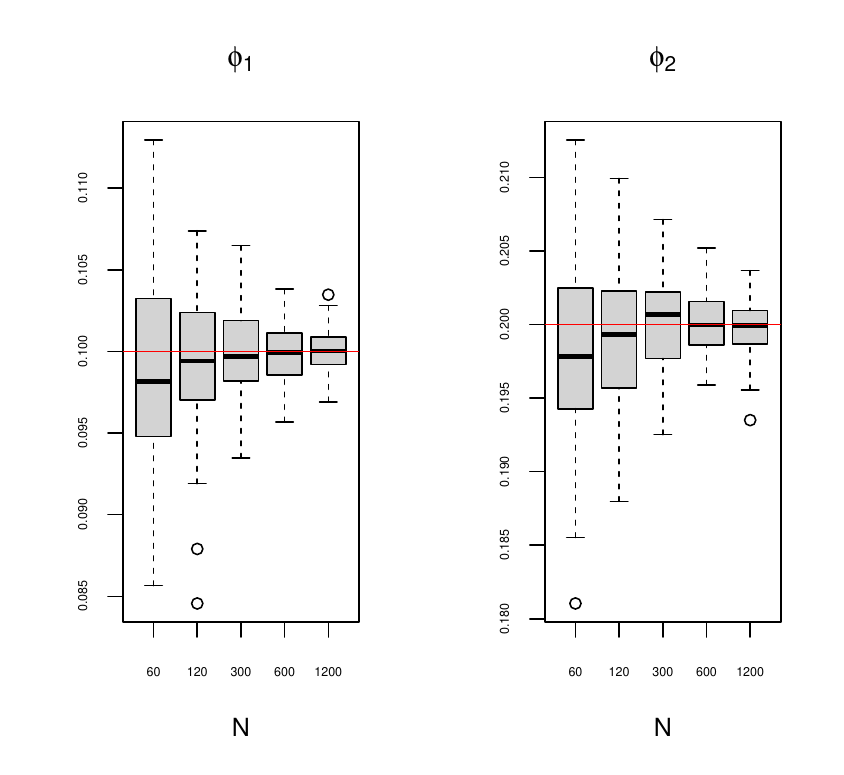}\\
\includegraphics[width=0.65\textwidth]{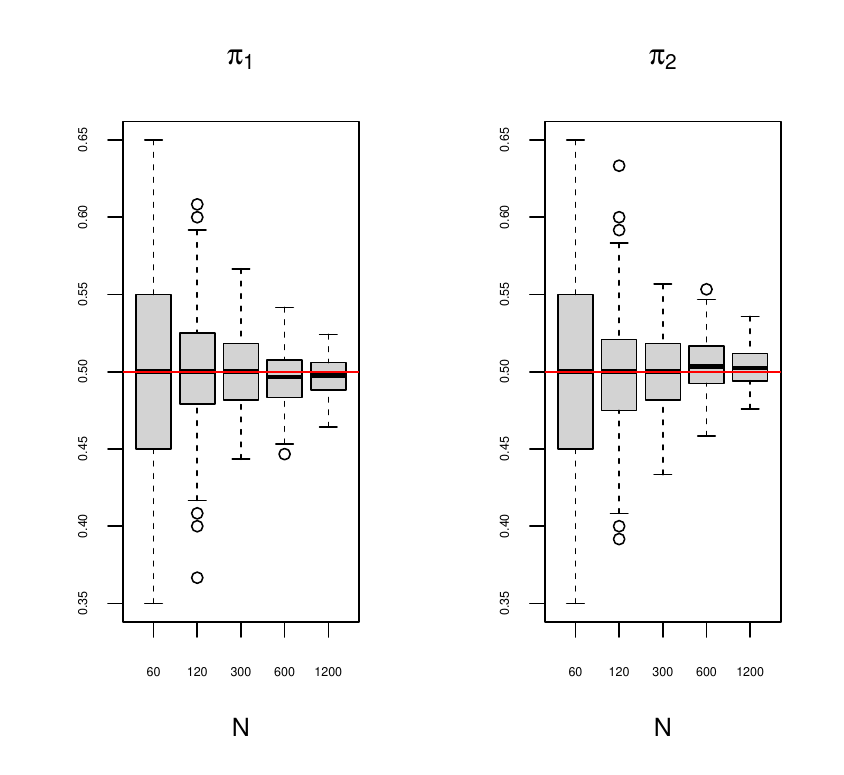}
\caption{\textbf{ZINB mixture model with a size factor:} Boxplots for the estimates of $\phi_k$ (\textit{top row}) and \(\pi_k\) (\textit{bottom row}) using the EM algorithm across the datasets simulated from the settings described in Table~\ref{t:SC1_ZINB2}. Red lines correspond to true values.
See also \textcolor{black}{Supplementary  Tables 89 
 and 90}.}
\label{f:SC1_ZINB2-phi-pi}
\end{figure}

\begin{figure}
\centering
\includegraphics[width=0.65\textwidth]{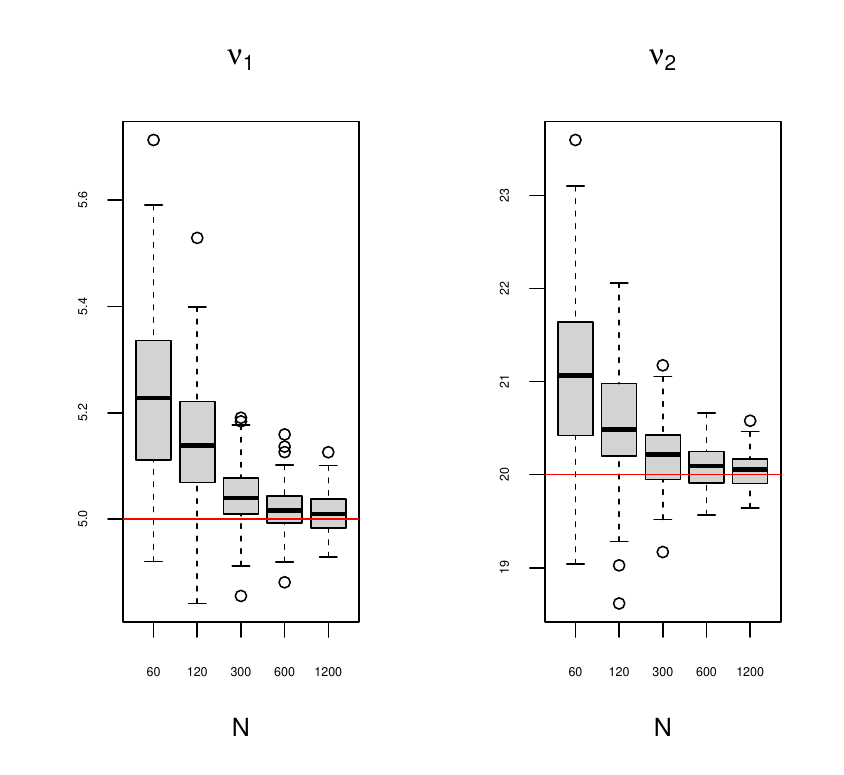}
\caption{\textbf{ZINB mixture model with a size factor:} Boxplots for the estimates of \(\nu_k\) using the EM algorithm across the datasets simulated from the settings described in Table~\ref{t:SC1_ZINB2}. Red lines correspond to true values.
See also \textcolor{black}{Supplementary Table 91 
}.}
\label{f:SC1_ZINB2-nu}
\end{figure}

In Figure \ref{f:SC1_ZINB2-phi-pi} and Supplementary Table 94,
we can see that the estimated cluster assignment probabilities ($\hat{\pi}_k$) are close to their true values, and their standard deviations decrease as $N$ increases. From Figure \ref{f:SC1_ZINB2-phi-pi} and \textcolor{black}{Supplementary Table 95},
we can observe similar behaviour for the estimated probabilities of always zero ($\hat{\phi}_k$); that is, as $N$ increases, their standard deviations decrease, and their values are close to the true values. Tables \ref{t:SC1_ZINB2-rho} and \ref{t:SC1_ZINB2-beta} present the MSEs for the estimated values of $\rho_{gk}$ and $\beta_{0g}$. We can observe that the MSEs reduce as $N$ becomes larger for both parameters. Figure \ref{f:SC1_ZINB2-nu} and Supplementary Table 96 
illustrate the behavior of the estimated size parameters ($\nu_1$ and $\nu_2$). As expected, the estimated values of $\nu_1$ and $\nu_2$ get closer to their true values, and their standard deviations decrease as the sample size ($N$) increases. In all cases, the V-measures are equal to one, indicating perfect clustering assignment results. Finally, \textcolor{black}{Supplementary Table 97 
and Supplementary Figure 52 
} show that the computing time increases as $N$ increases.

\begin{table}
\centering
\caption{\textbf{ZINB mixture model with a size factor:} Mean Squared error (MSE) for the estimates of $\rho_{gk}$ for each $k$ and each $N$, using the EM algorithm across the datasets simulated from the settings described in Table~\ref{t:SC1_ZINB2}.}
\label{t:SC1_ZINB2-rho}
\begin{tabular}{rrrrrr}
  \toprule
  & \multicolumn{5}{c}{$N$}\\
 \cmidrule(l){2-6}
    $k$&$60$ & $120$ & $300$ & $600$ & $1200$ \\
  \midrule
1 & 0.08710 & 0.06189 & 0.04038 & 0.02848 & 0.01996 \\ 
 2 & 0.04873 & 0.03532 & 0.02200 & 0.01585 & 0.01127 \\
   \bottomrule
\end{tabular}
\end{table}

\begin{table}
\centering
\caption{\textbf{ZINB mixture model with a size factor:} Mean squared error (MSE) for the estimates of $\beta_{0g}$ for each $N$, using the EM algorithm across the datasets simulated from the settings described in Table~\ref{t:SC1_ZINB2}.}
\label{t:SC1_ZINB2-beta}
\begin{tabular}{l*{5}{r}}
\toprule
$N$ & $60$ & $120$ & $300$ & $600$ & $1200$ \\ 
\midrule
$\beta_{0g}$ &  0.04207 & 0.02902 & 0.01791 & 0.01278 & 0.00935 \\
\bottomrule
\end{tabular}
\end{table}

\clearpage

\section{Data Analysis}\label{ch:data_analysis}
Zero-inflated count data often arise in single-cell RNA sequencing (scRNA-seq) experiments, where the high frequency of zeros (zero inflation) may result from either biological or technological noise. Therefore, in this section, we illustrate the utility of our proposed model-based clustering method introduced in Section \ref{sec:method} by applying it to two publicly available scRNA-seq datasets. In Section \ref{Data_1}, we consider scRNA-seq data from mouse embryonic stem cells collected by \cite{inDrop} and fit the ZIP mixture model without covariates as well as the ZIP mixture model with a size factor. In Section \ref{Data_2}, we analyze another scRNA-seq data from mouse liver tissue cells profiled by \cite{Tissues} using the ZIP mixture model without covariates and the ZINB mixture without covariates. 

\subsection{Mouse Embryonic Stem Cell (MESC) data}\label{Data_1}

\citeauthor{inDrop} \cite{inDrop} developed a laboratory platform (called inDrop from \underline{in}dexing \underline{Drop}lets) for indexing thousands of individual cells for RNA sequencing. \cite{inDrop} then used inDrop to obtain single-cell RNA sequencing data from mouse embryonic stem cells before (day 0) and after leukemia inhibitory factor (LIF) withdrawal (days 2, 4, and 7). Read counts across cells and genes for the different experiment days in \cite{inDrop} are publicly available through the Gene Expression Omnibus online repository under the accession code GSE65525. 

For our analysis, we consider the pooled data for day $0$ ($933$ cells) and day $4$ ($683$ cells) for a total of $N=\num{1616}$ cells (as subjects) and $G=\num{24175}$ genes (as observations). Then, as a common step in the analysis of scRNA-seq data \citep{BackSPIN2015,inDrop,SCRNA_seq_new1}, we filter out genes with very little variation across cells. In particular, we filter out genes with a read count interquartile range across cells smaller than one ($\mbox{IQR}=Q_3-Q_1\leq 1$), resulting in $\num{4514}$ genes initially selected. From these $\num{4514}$ genes, we select $100$ of them with the highest read count standard deviations across cells. Therefore, we continue the data analysis in this section using the read count data for $N=\num{1616}$ cells (subjects) and the selected $G=100$ most variable genes (observations). Note that our choice to select 100 genes was based on reducing the computation complexity of running different models for different number of clusters with different initializations. However, one can consider a larger number of genes and compare the results. 

We fit the mouse embryonic stem cell data (MESC) data considering the ZIP mixture model without covariates (Section \ref{ZIP_NOX}) and with a size factor (as the total number of read counts for each cell before performing any gene filtering) as shown in Equation (\ref{reduced_lambda}) of Section \ref{ZIP_X}. For each model, we apply the proposed EM algorithm considering different choices of $K$ (total number of clusters) and two clustering initialization methods: $K$-means and random clustering. After obtaining initial cluster assignments for the cells by $K$-means or random clustering, we can find the initial parameter values required to start the EM algorithm as follows. For the cluster probabilities, the $\pi_k$'s, we set their initial values to the proportion of cells assigned to each initial cluster. For the probabilities of always zero, the $\phi_k$'s, we set each $\phi_k^{(0)}$ to the proportion of zero entries in each cluster. For the case of the ZIP mixture model without covariates (simple ZIP model), for each initial cluster, we take the mean read count for each gene as the initial values of the $\lambda_{gk}$'s. For the ZIP mixture model with a size factor, we initialize the $\beta_{0g}$'s at zero, and the cluster effects, the $\rho_{gk}$'s, as the mean read count for each gene for each initial cluster.

For each choice of $K$ and each initialization method (random or $K$-means), we run the EM algorithm $32$ times corresponding to $32$ different initialization runs from different seeds. Next, for each initialization method, we choose the run with the smallest Aikaike Information Criterion (AIC) for each possible total number of clusters $K$. For each initialization method, after choosing the best run over each $K$, we use the elbow method to select the optimum number of clusters. The elbow takes the point with the highest AIC (on the $y$-axis) and the point with the highest $K$ (in the $x$-axis) and defines a line, usually going from the top-left to the bottom-right on the plot. The optimum point is then determined to be the one that is the farthest away below this line.

\subsubsection{Results of fitting the ZIP mixture model without covariates to the MESC data}\label{sec:simpleZIP_MESC}
This section presents the results of fitting the ZIP mixture model without covariates (simple ZIP, Section \ref{ZIP_NOX}) to the MESC dataset. As mentioned above, for our analysis, we use the pooled data of day $0$ and day $4$ over the $100$ most variable selected genes and $\num{1616}$ cells. 


\textcolor{black}{Supplementary Figures 53 and 54 
}
show the smallest AIC over the 32 EM runs for each $K$ when considering random and $K$-means initialization methods, respectively. Based on the elbow method, both initialization methods lead to $K=4$ as the best number of clusters for the simple ZIP model.  
Next, as a final choice between these two results with $K=4$, one using random clustering initialization and the other $K$-means, we select the one with the lowest AIC. In this case, the AIC from the $K$-means  initialization approach reaches a lower value than that of the random initialization. Therefore, we select the EM run with the best AIC for $K=4$ from the $K$-means initialization approach and present its results in what follows.

\begin{figure}
\centering
\includegraphics[width=0.65\textwidth]{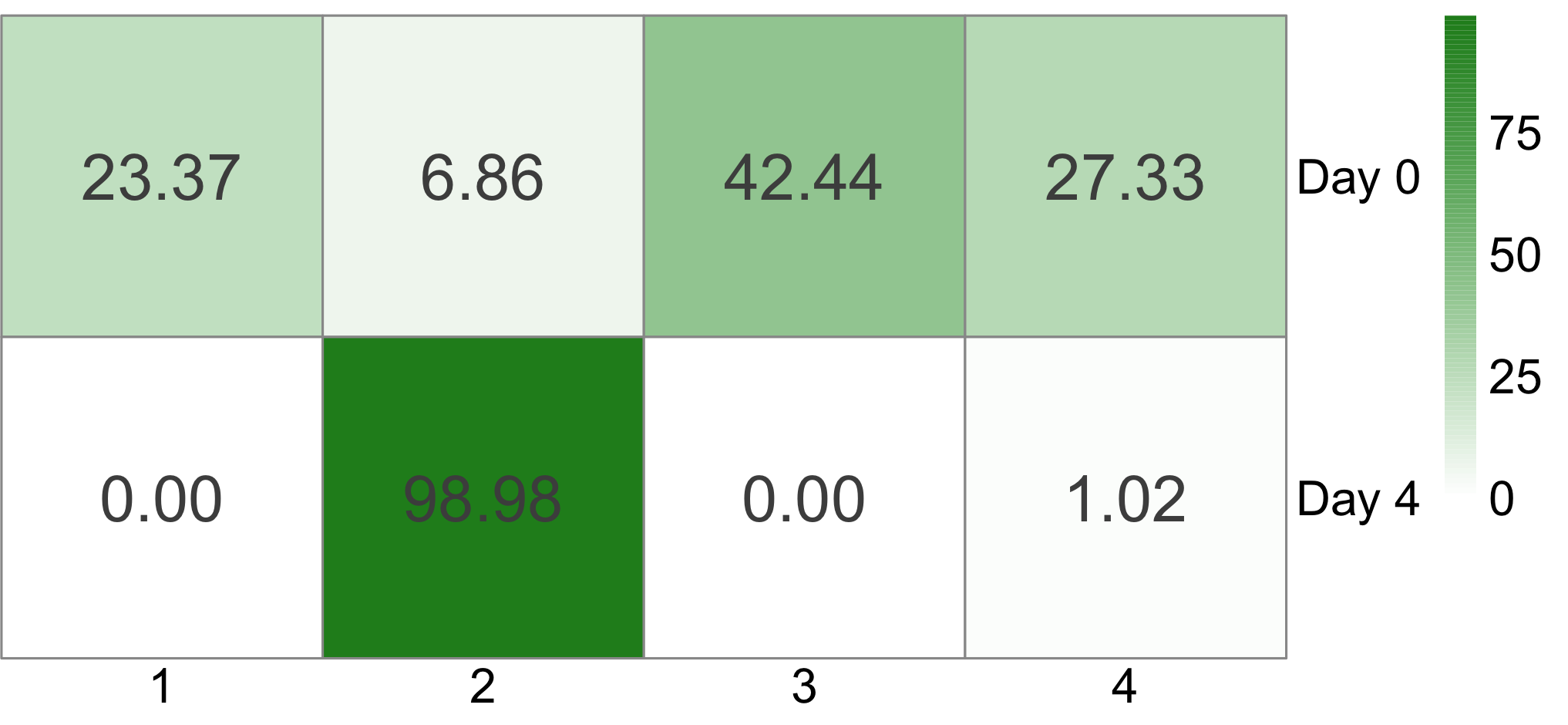}
\caption{MESC dataset. Co-clustering between experiment days (0 and 4; rows)  and inferred clusters by the proposed EM algorithm (1, 2, 3, and 4; columns). Each entry $a_{ij}$ represents the $\%$ of cells from day $i$ that are present in the inferred cluster $j$. Rows sum up to $100\%$. Inferred clusters are from the best EM algorithm run ($K$-means initialization and $K=4$) under the simple ZIP model.}
\label{f:v6m1-6}
\end{figure}

Figure \ref{f:v6m1-6} and \textcolor{black}{Supplementary Table 98 
} present the co-clustering plot and confusion matrix, respectively, between cell experiment days (day 0 and day 4) and the inferred cell clusters (1, 2, 3, and 4) for the best EM algorithm run with $K=4$. The co-clustering plot allows us to observe the percentage of cells from each experiment day present in each inferred cluster. We can see that most cells (approximately $99\%$) of day $4$ are present in the inferred cluster $2$ and cells from day $0$ fall mainly into the inferred clusters $1$, $3$, and $4$. Interestingly, \cite{inDrop} found in their analyses that cells from day $0$ belong to three main subpopulations plus two other rare subpopulations when clustering only day $0$ cells via hierarchical clustering.

\begin{figure}
\centering
\includegraphics[width=\textwidth]{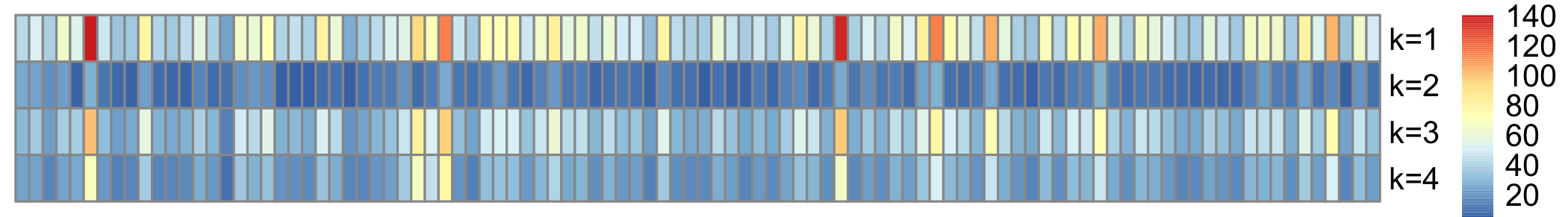}
\caption[Heatmap of the $\hat{\lambda}_{gk}$'s for the MESC dataset obtained from the best EM algorithm run ($K$-means initialization and $K=4$) under the simple ZIP model.]{Heatmap of the $\hat{\lambda}_{gk}$'s for MESC data set obtained from the best EM algorithm run ($K$-means initialization and $K=4$) under the simple ZIP model. Each row corresponds to a cluster, and each column to a gene. Dark blue colours represent low values, and dark red colours represent high values of $\hat{\lambda}_{gk}$.}
\label{f:v6m1-9}
\end{figure}

Table \ref{t:v6m1-pi-phi} shows the estimated cluster proportion, $\hat{\pi}_k$, for each inferred cluster. Cluster $2$ has the highest proportion of cell assignments ($\hat{\pi}_2=45.91\%$) compared to the other three clusters. The estimated probability of always zero for each cluster is also displayed in Table \ref{t:v6m1-pi-phi} with cluster $2$ showing the highest probability at $\hat{\phi}_2=1.191\%$. Figure \ref{f:v6m1-9} shows the heatmap of the rate parameter estimates ($\hat{\lambda}_{gk}$'s) for each cluster (rows) over the $100$ genes (columns) when fitting the ZIP simple model to the MESC data. 

\textcolor{black}{Supplementary Figure 55 
} shows the heatmap of the data (read counts across all $\num{1616}$ cells and all $100$ selected genes) with cells (rows) ordered by their inferred cluster assignments. The heatmap also contains annotation for each cell's experiment day (0 or 4). \textcolor{black}{Supplementary Figure 56 
} presents a dimensionality reduction visualization of the data using $t$-SNE \citep{tsne_van2008visualizing}. In the $t$-SNE plot, circle and triangle points correspond to cells from day 0 and day 4, respectively, and the colours to the inferred four clusters. Moreover, \textcolor{black}{Supplementary Figure 57 
} shows the cluster assignment expected values (or probabilities), i.e.,  the $\hat{Z}_{nk}$'s, which we used to determine the final inferred cluster assignment of each cell. 
Overall, we can observe that the proposed EM algorithm assigned cells to their clusters with high (close to 1) probabilities.

\begin{table}
\centering
\caption{Estimates of $\pi_k$ and $\phi_k$ for the MESC dataset obtained from the best EM algorithm run ($K$-means initialization and $K=4$) under the simple ZIP model.} 
\label{t:v6m1-pi-phi}
\begin{tabular}{lrr}
  \toprule
  $k$ & $\hat\pi_k$&$\hat\phi_k$\\
 \midrule
1 & 0.13599&  0.00077 \\ 
  2 & 0.45910& 0.01191\\ 
  3 & 0.24507& 0.00077\\ 
  4 & 0.15984& 0.00127\\ 
   \bottomrule
\end{tabular}
\end{table}






\subsubsection{Results of fitting the ZIP mixture model with a size factor to the MESC data}\label{sec:ZIPfactor_MESC}
In this section, we present the results of fitting the ZIP mixture model with a size factor (as shown in Equation (\ref{reduced_lambda}) of Section \ref{ZIP_X}) to the same MESC dataset analyzed in Section \ref{sec:simpleZIP_MESC}; that is, the pooled data of day $0$ and day $4$ over the $100$ most variable selected genes and $\num{1616}$ cells. 


\textcolor{black}{Supplementary Figures 58 and 59 
} show the smallest AIC for each $K$ for random and $K$-means initialization methods, respectively. Based on the elbow method, the random initialization approach leads to $K=6$ clusters as the best number of clusters for the ZIP mixture model with size factor; however, the $K$-means initialization method leads to $K=4$ as the optimum number for clusters.
As a final choice, we choose the best number of clusters between these two initialization methods based on the lowest AIC. Thus, as $K$-means initialization leads to the smallest AIC value, we choose the EM run with the best AIC from the $K$-means method when $K=4$ and present its results in the following. 

\begin{figure}
\centering
\includegraphics[width=0.65\textwidth]{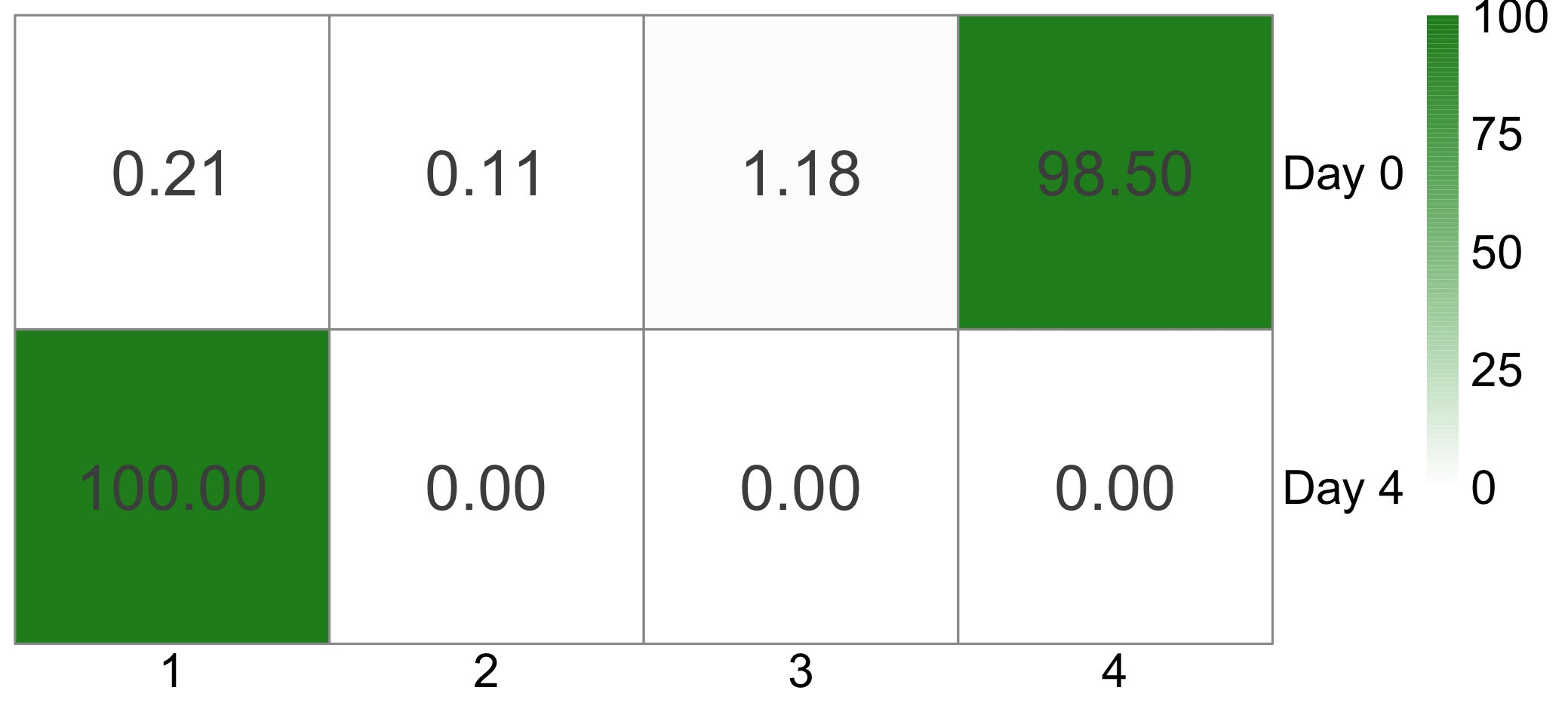}
\caption{MESC dataset. Co-clustering between experiment days (0 and 4; rows)  and inferred clusters by the proposed EM algorithm (1, 2, 3, and 4; columns). Each entry $a_{ij}$ represents the $\%$ of cells from day $i$ that are present in the inferred cluster $j$. Rows sum up to $100\%$. Inferred clusters are from the best EM algorithm run ($K$-means initialization and $K=4$) under the ZIP mixture model with a size factor.}
\label{f:v6m2-6}
\end{figure}

Figure \ref{f:v6m2-6} and \textcolor{black}{Supplementary Table 99 
} display the co-clustering plot and confusion matrix, respectively, between cell experiment days ($0$ and $4$) and the inferred four clusters from the best EM algorithm run ($K=4$ clusters and $K$-means clustering initialization method) under a ZIP mixture model with a size factor. Notably, all cells ($100\%$) from day $4$ are present in the inferred cluster $1$, while most cells (approximately $98.5\%$) from day $0$ are in the inferred cluster $4$. Only a small portion of cells from day $0$ are found in the other inferred clusters: $0.21\%$ in cluster $1$, $0.11\%$ in cluster $2$, and $1.18\%$ in cluster $3$. Interestingly, these clustering results are similar to the ones presented in \cite{DAYS}.

\begin{figure}
\centering
\includegraphics[width=\textwidth]{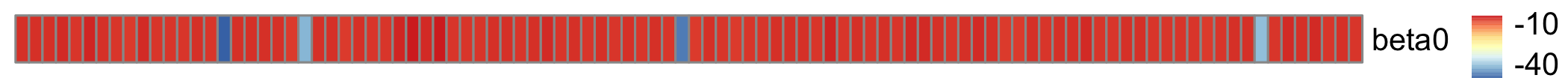}
\includegraphics[width=\textwidth]{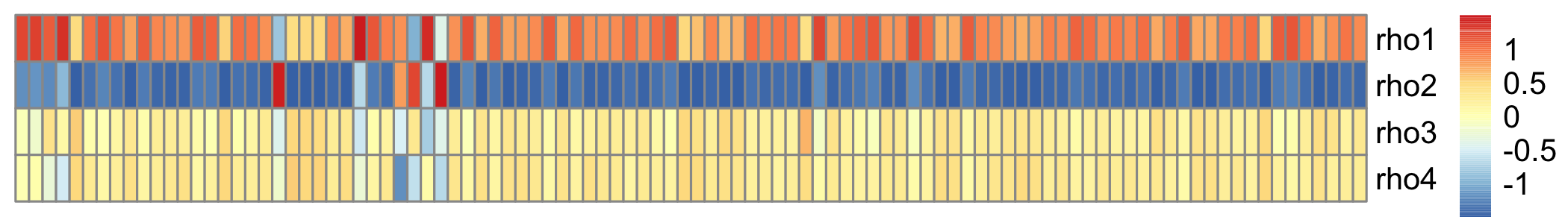}
\caption[Heatmap of the $\hat{\beta}_{0g}$'s and $\hat{\rho}_{gk}$'s for the MESC dataset obtained from the best EM algorithm run ($K$-means initialization and $K=4$) under the ZIP mixture model with size factor.]{Heatmap of the $\hat{\beta}_{0g}$'s and $\hat{\lambda}_{gk}$'s for the MESC dataset obtained from the best EM algorithm run ($K$-means initialization and $K=4$) under the ZIP mixture model with size factor. $\hat{\beta}_{0g}$'s are shown in the first row, and the $\hat{\rho}_{gk}$'s for each cluster $k$ are presented in rows $2$ to $5$. The columns correspond to the 100 selected genes. Dark blue colours represent low values, and dark red colours represent high values of $\hat{\beta}_{0g}$ and $\hat{\rho}_{gk}$.}
\label{f:v6m2-9}
\end{figure}

The estimated cluster proportions ($\hat{\pi}_k$'s) are presented in Table \ref{t:v6m2-pi-phi}. We can observe that more than $50 \%$ of the cells belong to cluster $4$ ($56.83\%$), and $42.4\%$ of the cells fall into cluster $1$ and only a few of them are assigned to the other two clusters. Table \ref{t:v6m2-pi-phi} also shows the estimated probability of always zero for each cluster $k$, and we can see that $\hat{\phi}_1=0.01415$ has the highest probability of always zero compared with the other clusters.
Figure \ref{f:v6m2-9} 
shows the heatmap of the estimates of $\beta_{0g}$ (baseline expression) and $\rho_{gk}$ (cluster effect) over the $100$ genes (columns) when fitting the ZIP mixture model with a size factor. The $\hat{\beta}_{0g}$'s are shown in the first row, and the $\hat{\rho}_{gk}$'s for each cluster $k$ are presented in rows $2$ to $5$.

Similarly to the previous section, \textcolor{black}{Supplementary Figure 60 
} shows the heatmap of the read counts across all $\num{1616}$ cells and all $100$ selected genes with cells (rows) ordered by their inferred cluster assignments from the ZIP mixture model with size factor. The heatmap also shows each cell's experiment day ($0$ or $4$).
\textcolor{black}{Supplementary Figure 61 
} shows the $t$-SNE representation of the data in two dimensions, where circle and triangle points correspond to cells from day 0 and day 4, respectively, and the colours to the inferred four clusters. As in Section \ref{sec:simpleZIP_MESC}, we can observe in \textcolor{black}{Supplementary Figure 62 
} that overall the proposed EM algorithm assigned cells to their clusters with high (close to $1$) probabilities.






\begin{table}
\centering
\caption{Estimates of $\pi_k$ and $\phi_k$ for the MESC dataset obtained from the best EM algorithm run ($K$-means initialization and $K=4$) under the ZIP mixture model with size factor.} 
\label{t:v6m2-pi-phi}
\begin{tabular}{lrr}
  \toprule
  $k$ & $\hat\pi_k$& $\hat\phi_k$\\
 \midrule
1 & 0.42384 &0.01415\\ 
  2 & 0.00062& 0.00000\\ 
  3 & 0.00727& 0.00086\\ 
  4 & 0.56828& 0.00072\\ 
   \bottomrule
\end{tabular}
\end{table}

\subsubsection{Model selection for the MESC dataset}
For the best EM algorithm runs when fitting the simple ZIP model (ZIP mixture model without covariates), we obtained the AIC value of $\num{80274.22}$ and the AIC for the ZIP mixture model with a size factor was $\num{72544.74}$   
These values correspond to the red points in \textcolor{black}{Supplementary Figures 54 and 59 
}. 
The resulted AIC values show that the ZIP mixture model with a size factor leads to a smaller AIC value than the simple ZIP model. Therefore, the ZIP mixture model with a size factor fits the MESC data better than the ZIP simple model. As shown in Section \ref{sec:ZIPfactor_MESC}, the selected ZIP mixture model with size factor resulted in two main clusters, one with cells from day 0 and the other with cells from day 4. 


\subsection{Liver Data}\label{Data_2}
\citeauthor{Tissues} \cite{Tissues} collected thousands of single-cell transcriptome profiles from several mouse tissues, organs, and cell cultures using Microwell-seq as a high-throughput and low-cost scRNA-seq platform.
The data is publicly available at \url{https://figshare.com/s/865e694ad06d5857db4b} and in the Gene Expression Omnibus online repository under the accession code GSE108097. In this section, we focus on a subset of the scRNA-seq data from mouse liver tissue provided by \cite{Tissues}, which comprises a total of $N=\num{4685}$ cells and $G=\num{15491}$ genes. We first select $N=\num{1000}$ cells from the total $\num{4685}$ cells using random sampling. Then, as in Section \ref{Data_1}, we filter out genes with minimal variation across these randomly selected cells, keeping only a set of highly variable genes. In particular, we choose $100$ genes with the highest standard deviations of read count across cells. Therefore, the data analysis results presented in this section are based on the randomly selected $\num{1000}$ cells and their corresponding $100$ highly variable genes of the liver tissue data in \cite{Tissues}, which will be referred to as the liver data hereafter.

We fit the liver data considering the ZIP mixture model without covariates (simple ZIP, Section \ref{ZIP_NOX}) and the ZINB mixture model without covariates (simple ZINB, Section \ref{ZINB_NOX}) via the EM algorithm. Similarly to Section \ref{Data_1}, we apply the proposed EM algorithm to the liver data considering different choices of $K$ (total number of clusters) and two clustering initialization approaches: $K$-means and random clustering. After obtaining the initial cluster assignments for the cells using the two initialization methods, we can find the initial starting points of the EM algorithm for each model fitting. 
We can obtain the starting points for the cluster probabilities ($\pi_k$'s), proportions of always zero in each cluster ($\phi_k$'s) for both the simple ZIP and ZINB mixture models and the rate parameters of simple ZIP model as described earlier in Section \ref{Data_1}. 
The initial rate parameters for each cluster under the simple ZINB mixture model are the mean read counts for each \textcolor{black}{gene across cells} in each initial cluster (similar to the initial rates for the simple ZIP mixture model). Finally, we calculate the initial values for the size parameters ($\nu_k$'s) for the simple ZINB mixture model. For each $k$, first, we calculate $\mu_k$ and $\sigma_k$ as the mean and standard deviation of \textcolor{black}{read counts over all genes and cells} in the initial cluster $k$. Then, we calculate the initial value for $\nu_k$ as:
$$\nu_k^{(0)}=\Big[\Big(\frac{\sigma_k}{\mu_k}\Big)^2-\frac{1}{\mu_k}\Big]^{-1}.$$

Again, similar to Section \ref{Data_1}, for each choice of $K$ and initialization method, we run the proposed EM algorithm $32$ times as $32$ different initialization runs from different seeds and choose the run with the smallest AIC for each possible $K$ for each initialization method. Then, after selecting the best run for each $K$, we use the elbow method to find the optimum number of clusters $K$.

\subsubsection{Result of fitting the ZIP mixture model without covariates to the liver data}\label{sec:ZIP_liver}
This section presents the results of fitting the ZIP mixture model (simple ZIP) to the liver data set. The data correspond to the  $\num{1000}$ randomly selected cells and the $100$ selected highly variable genes for our analysis.

Figures 63 and 64 in the supplementary 
show the smallest AIC for each $K$ for random and $K$-means initialization methods, respectively. Based on the elbow method, the random initialization approach leads to $K=6$ clusters as the best number of clusters. The $K$-means approach yields $K=4$ as the optimum number of clusters. As the best EM run from $K$-means initialization has the lowest value of AIC, in what follows, we present the results of the EM run with $K=4$ from the $K$-means initialization method.  

Table \ref{t:v21m1-confmat} and Supplementary Figure 65 
show the confusion matrix and co-clustering plot, respectively, between the cell type labels provided in the dataset by \cite{Tissues} and the inferred four clusters for the best EM run ($K=4$ clusters and $K$-means initialization approach) for the ZIP mixture model without covariates. We can observe that B cells ($62.16\%$) are mostly assigned to the inferred cluster $3$, dendritic cells are present equally in clusters $1$ ($50\%$) and $3$ ($50\%$), endothelial cells ($91.76\%$) are mainly assigned to cluster $3$, all epithelial cells are part of the inferred cluster $2$, $77.59\%$ of erythroblast cells and $84.62\%$ of granulocyte cells are assigned to cluster $3$, $90.14\%$ of hepatocyte cells are in the inferred cluster $2$, $67.14\%$ and $32.86\%$ of Kuppfer cells are assigned to clusters $1$ and $3$, respectively. $58\%$ of the macrophage cells are present in cluster $3$, and $42\%$ of them are in cluster $1$. All ($100\%$) of the neutrophil cells and $92.65\%$ of the T cells are assigned to the inferred cluster $3$.

\begin{table}
\centering
\caption{Confusion matrix between the EM clustering result when fitting the simple ZIP model to the liver data and the cell types. Inferred clusters are from the best EM algorithm run ($K$-means initialization and $K=4$) under the simple ZIP model (see also Supplementary Figure 62).}
\label{t:v21m1-confmat}
\begin{tabular}{lrrrr}
  \toprule
  & \multicolumn{4}{c}{Inferred cluster}\\
 \cmidrule(l){2-5}
Cell type & 1 & 2 & 3 & 4 \\
\midrule
B cell           &   3 &  1 &  23 & 10 \\
Dendritic cell   &  54 &  0 &  54 &  0 \\
Endothelial cell &  17 &  5 & 245 &  0 \\
Epithelial cell  &   0 & 32 &   0 &  0 \\
Erythroblast     &   9 & 17 &  90 &  0 \\
Granulocyte      &   5 &  1 &  33 &  0 \\
Hepatocyte       &   2 & 64 &   5 &  0 \\
Kuppfer cell     & 141 &  0 &  69 &  0 \\
Macrophage       &  21 &  0 &  29 &  0 \\
Neutrophil       &   0 &  0 &   2 &  0 \\
T cell           &   5 &  0 &  63 &  0 \\
   \bottomrule
\end{tabular}
\end{table}

\begin{figure}
\centering
\includegraphics[width=\textwidth]{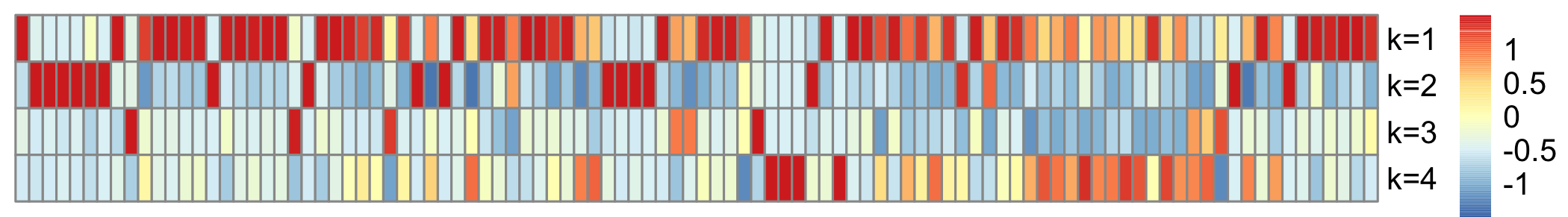}
\caption[Heatmap of the $\hat{\lambda}_{gk}$'s for the liver dataset obtained from the best EM algorithm run ($K$-means initialization and $K=4$) under the simple ZIP model.]{Heatmap of the $\hat{\lambda}_{gk}$'s for the liver dataset obtained from the best EM algorithm run ($K$-means initialization and $K=4$) under the simple ZIP model. Each row corresponds to a cluster, and each column to a gene. Dark blue colours represent low values, and dark red colours represent high values of $\hat{\lambda}_{gk}$.}
\label{f:v21-m1-9}
\end{figure}

The estimated cluster proportions ($\hat{\pi}_k$'s) and probabilities of always zero ($\hat{\phi}_k$'s) are presented in Table \ref{t:v21m1-pi-phi}. The table shows that $60.929\%$ of the cells are assigned in the inferred cluster $3$, $26.114\%$ are assigned to cluster $1$, $11.957\%$ fall into cluster $2$, and only $1\%$ of them fall into cluster $4$. We can also see that the higher estimated probabilities of always zero are for the inferred clusters $2$ and $4$, $\hat{\phi}_2=0.26895$ and $\hat{\phi}_4=0.13620$, respectively, followed by clusters $1$ and $3$ ($\hat{\phi}_1=0.09183$ and $\hat{\phi}_3=0.08780$). 
Finally, Figure \ref{f:v21-m1-9} shows the heatmap of the estimates of the rate parameters ($\hat{\lambda}_{gk}$'s) for each cluster (rows) over the $100$ selected genes (columns) when fitting the ZIP simple model to the liver data.

Figure 66 in the Supplementary Material
displays the heatmap of the \textcolor{black}{read counts across the randomly selected $1000$ cells and all $100$ selected genes}. The cells (rows) are ordered by their inferred cluster assignments from the simple ZIP mixture model. The heatmap also shows each cell's type. Additionally, Supplementary Figure 67 
shows the $t$-SNE representation of the data in two dimensions, where different point shapes correspond to the cell type and point colours to the inferred four clusters. As in Sections \ref{sec:simpleZIP_MESC} and \ref{sec:ZIPfactor_MESC}, Supplementary Figure 68 
demonstrates that the proposed EM algorithm assigned cells to their clusters with high (close to $1$) probabilities.





\begin{table}
\centering
\caption{Estimates of $\pi_k$ and $\phi_k$ for the liver dataset obtained from the best EM algorithm run ($K$-means initialization and $K=4$) under the simple ZIP model.}
\label{t:v21m1-pi-phi}
\begin{tabular}{lrr}
  \toprule
  $k$ & $\hat\pi_k$& $\hat\phi_k$\\
 \midrule
1 & 0.26114& 0.09183\\ 
  2 & 0.11957& 0.26895\\ 
  3 & 0.60929 &0.08780\\ 
  4 & 0.01000 &0.13620\\ 
   \bottomrule
\end{tabular}
\end{table}

\subsubsection{Result of fitting the ZINB mixture model without covariates to the liver data}\label{sec:ZINB_liver}
In this section, the results of fitting the ZINB mixture model without covariates (simple ZINB) for the liver data (over the selected $1000$ cells and $100$ selected mostly variable genes, same data as in Section \ref{sec:ZIP_liver}) are presented. 

Supplementary Figures 69 and 70 
present the corresponding smallest AIC for each $K$ for the random and $K$-means initialization methods, respectively. Based on the elbow method, random initialization results in $K=4$ as the optimum number of clusters, while $K$-means leads to $K=5$ as the best number of clusters. As the one from $K$-means clustering initialization has the smallest AIC, we choose that EM run for further analysis in this section. 
Therefore, for the simple ZINB mixture model, we choose the EM run with $K=5$ from the $K$-means initialization method, and we present its results in what follows.  

Table \ref{t:v21m3-confmat} and Supplementary Figure 71 
show the confusion matrix and co-clustering plot, respectively, between the cell types and the inferred five clusters for the best EM run ($K=5$ clusters and $K$-means initialization approach) for the ZINB mixture model without covariates. We can see that B cells are assigned primarily to $3$ clusters, with $43.24\%$ of them in cluster $3$, $35.14\%$ in cluster $5$, and $13.51\%$ in cluster $2$.
Dendritic cells are primarily present in clusters $2$ ($85.19\%$), endothelial cells ($91.39\%$) are dense in cluster $3$, epithelial ($90.62\%$) cells are mainly in the inferred cluster $1$. $75\%$ of erythroblast cells and $69.23\%$ of Granulocyte cells fall into cluster $3$. $84.51\%$ of Hepatocyte cells are in the inferred cluster $1$, Kuppfer cells are mostly assigned to clusters $3$ and $4$, with $19.05\%$ and $80\%$, respectively. $70\%$ of macrophage cells fall into cluster $2$ and $22\%$ of them are in cluster $3$, and the remaining $8\%$ in cluster $4$.
Neutrophil cells are assigned equally ($50\%$) to the inferred clusters $3$ and $5$. Finally, $77.94\%$ of T cells fall into the inferred cluster $3$.

\begin{table}
\centering
\caption{Confusion matrix between the EM clustering result when fitting the simple ZINB model to the liver data and the cell type labels. Inferred clusters are from the best EM algorithm run ($K$-means initialization and $K=5$) under the simple ZINB model (see also Supplementary Figure 68).}
\label{t:v21m3-confmat}
\begin{tabular}{lrrrrr}
\toprule
& \multicolumn{5}{c}{Inferred cluster} \\
\cmidrule(l){2-6}
Cell type & 1 & 2 & 3 & 4 & 5 \\
\midrule
B cell           &  0 &  5 &  16 &   3 & 13 \\
Dendritic cell   &  0 & 92 &  12 &   3 &  1 \\
Endothelial cell &  0 &  3 & 244 &  17 &  3 \\
Epithelial cell  & 29 &  0 &   0 &   0 &  3 \\
Erythroblast     &  1 &  3 &  87 &   9 & 16 \\
Granulocyte      &  0 &  3 &  27 &   3 &  6 \\
Hepatocyte       & 60 &  1 &   1 &   0 &  9 \\
Kuppfer cell     &  0 &  2 &  40 & 168 &  0 \\
Macrophage       &  0 & 35 &  11 &   4 &  0 \\
Neutrophil       &  0 &  0 &   1 &   0 &  1 \\
T cell           &  2 &  9 &  53 &   2 &  2 \\
\bottomrule
\end{tabular}
\end{table}

\begin{figure}
\centering
\includegraphics[width=\textwidth]{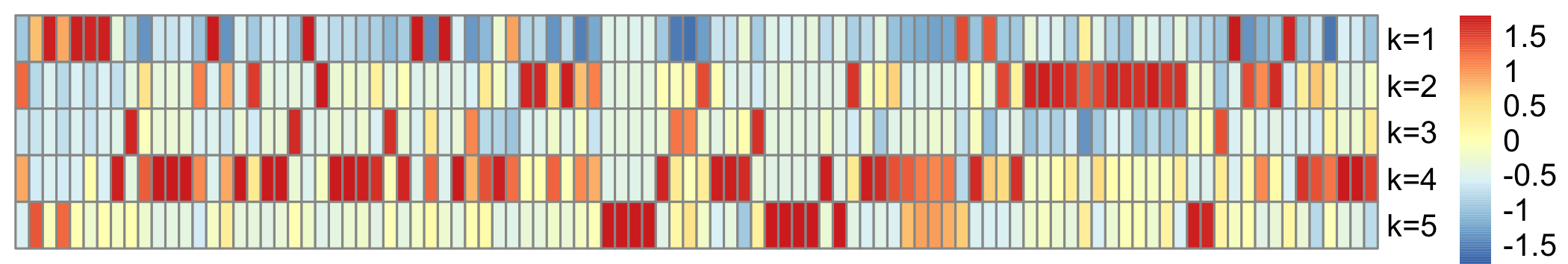}
\caption[Heatmap of the $\hat{\mu}_{gk}$'s for the liver dataset obtained from the best EM algorithm run ($K$-means initialization and $K=5$) under the simple ZINB model.]{Heatmap of the $\hat{\mu}_{gk}$'s for the liver dataset obtained from the best EM algorithm run ($K$-means initialization and $K=5$) under the simple ZINB model. Each row corresponds to a cluster, and each column to a gene. Dark blue colours represent low values, and dark red colours represent high values of $\hat{\mu}_{gk}$.}
\label{f:v21-m3-9}
\end{figure}

The estimated cluster proportions ($\hat{\pi}_k$'s) are presented in Table \ref{t:v21m3-pi-phi-nu}. We can see that $49.157\%$ of the \textcolor{black}{cells} are assigned to the inferred cluster $3$, $20.986\%$ are assigned to cluster $4$, $15.235\%$ fall into cluster $2$, $9.163\%$ are in cluster $1$, and only $5.54\%$ of them are assigned to cluster $5$.
Table \ref{t:v21m3-pi-phi-nu} also shows the estimated probability of always zero for each cluster $k$, and we can see that $\hat{\phi}_1=0.07266$ is the highest probability of always zero compared with the other clusters.
The estimates of the size parameters for each cluster ($\nu_k$'s) are also shown in Table \ref{t:v21m3-pi-phi-nu}. We can see that clusters $3$, $4$, and $2$ have the higher estimated values ($\hat{\nu}_3=4.11696,\,\hat{\nu}_4=3.75589,\,\hat{\nu}_2=2.12675$), followed by the smaller estimated values for the other two clusters ($\hat{\nu}_1=1.93291,\,\hat{\nu}_5=0.66367$), demonstrating the presence of overdispersion in the liver data.
Finally, Figure \ref{f:v21-m3-9} shows the heatmap of the estimates of the rate parameters ($\hat{\mu}_{gk}$'s) for each cluster (rows) over the $100$ selected genes (columns) when fitting the ZINB simple model to the liver data.

Supplementary Figure 72
shows the heatmap of the read counts across the randomly selected $1000$ cells and all $100$ most variable selected genes with cells (rows) ordered by their inferred cluster assignment from the simple ZINB mixture model. The heatmap also shows the annotation for each cell's type. Figure 73 in the Supplementary Material
shows the $t$-SNE representation of the data in two dimensions, where different point shapes correspond to the cell type and the colors to the inferred five clusters. As in the previous sections, we can observe in Supplementary Figure 74
that overall the proposed EM algorithm assigned cells to their clusters with high (close to $1$) probabilities.



\begin{table}
\centering
\caption{Estimates of $\pi_k$ for the liver dataset obtained from the best EM algorithm run ($K$-means initialization and $K=5$) under the simple ZINB model.}
\label{t:v21m3-pi-phi-nu}
\begin{tabular}{lrrr}
  \toprule
  $k$ & $\hat\pi_k$ &$\hat\phi_k$ & $\hat\nu_k$\\
 \midrule
1 & 0.09163 & 0.07266 & 1.93291\\ 
  2 & 0.15235 & 0.00000 &2.12675\\ 
  3 & 0.49157 &0.00005 &4.11696\\ 
  4 & 0.20986 &0.00000 &3.75589\\ 
  5 & 0.05458 &0.00007 &0.66367\\ 
   \bottomrule
\end{tabular}
\end{table}
\subsubsection{Model selection for the liver data}
For liver data, the AIC values obtained from the best EM algorithm runs when fitting the simple ZIP model (ZIP mixture model without covariates) was $\num{115744.1}$ and for the simple ZINB model (ZINB mixture model without covariates) equals $\num{360931}$.
These values correspond to the red points in Figures 64 and 70 in the Supplementary Material.
Thus, based on the AIC values presented above,
the simple ZINB model leads to a smaller AIC value than the simple ZIP model and, therefore, we can conclude that the simple ZINB fits the liver data better than the ZIP simple model. As shown in Section \ref{sec:ZINB_liver}, the selected simple ZINB model resulted in five clusters comprising different cell types in the liver tissue.
\section{Conclusion}\label{ch:conclusion}

In this paper, we introduced a model-based approach for clustering zero-inflated counts. We considered a mixture of zero-inflated Poisson (ZIP) or zero-inflated negative binomial (ZINB) distributions. Our method employs an EM algorithm to estimate model parameters and assign clusters, considering scenarios with and without covariates. Additionally, we incorporated a size factor to account for differences in total sampling or exposure, which can influence the observed counts. This framework effectively addresses the challenge of excess zeros while capturing underlying heterogeneity, making a significant contribution to model-based clustering of zero-inflated count data.


In Section \ref{Results} of our paper, we conducted simulation studies to assess the frequentist properties (bias and variance) of our proposed parameter estimators under various scenarios for each proposed mixture model. While some simulation results are included in Section \ref{Results}, additional scenarios and their corresponding results can be found in the Supplementary Material. The simulation studies demonstrated that as the number of subjects ($N$) increased, the bias, standard deviation, and Mean Squared Error (MSE) or Mean Absolute Deviation (MAD) decreased, consistent with the convergence properties of the EM algorithm. When the number of observations ($G$) increased, our proposed EM framework performed well in terms of bias, as almost all the estimated parameters were close to their true values. The standard deviations and MSEs (or MADs) for the estimated values of the cluster assignment probabilities and the rate parameters (for all models) remained almost the same when $G$ increased. However, when estimating the probability of always zero, we observed a decrease in standard deviation when increasing $G$. Furthermore, when increasing model complexity by increasing the number of clusters $K$, almost for all cases, the standard deviations, MSE, or MAD increased. At the same time, all the estimates remained close to their true values. 

In Section \ref{ch:data_analysis}, we applied the proposed models and developed EM algorithms to the MESC and liver scRNA-seq datasets. For the MESC data, we considered and compared the results from the ZIP mixture models without covariates and with a size factor. Using the AIC criterion for comparing the models, we selected the ZIP mixture model with a size factor as the final choice, resulting in $K=4$ clusters, which clustered \textcolor{black}{cells} mainly in $2$ clusters (all cells from experiment day $4$ were assigned to cluster $1$ and $98.5\%$ of the cells from experiment day $0$ were assigned to cluster $4$). Next, we fitted the simple ZIP and ZINB models (i.e., without covariates or size factors) to the liver tissue data. Comparing the AIC from the final fits between the simple ZIP and ZINB models led to the simple ZINB model with $K=5$ clusters as the model that best fitted the liver data. One of the challenges in analyzing scRNA-seq data is the selection/filtering of genes before model fitting. A sensitivity analysis considering different filtering methods available in the literature could be further investigated in future work.

Another future research opportunity regarding the proposed clustering methodology includes taking a Bayesian inference approach via MCMC or variational Bayes methods.

\section*{Acknowledgments}

This research is supported by the Natural Sciences and Engineering Research Council of Canada (NSERC).

\section*{Conflict of interest}

The authors declare that they have no conflict of interest.

\clearpage

\section*{Appendix}

\begin{algorithm}[htp]
  \caption{EM algorithm for the ZIP mixture model without covariates}
  \label{alg::ZIP_No_X}
  \begin{algorithmic}[1]
    \Require
      $\boldsymbol{y}$: matrix of data;
      $\boldsymbol{\theta}^{(0)}=(\boldsymbol{\pi}^{(0)},\boldsymbol{\phi}^{(0)},\boldsymbol{\lambda}^{(0)})$: initial parameters;
      $tol$: tolerance;
      $m$: maximum number of iterations.
    \Ensure
      optimal set of parameters $\hat{\boldsymbol{\theta}}=(\hat{\boldsymbol{\pi}},\hat{\boldsymbol{\phi}},\hat{\boldsymbol{\lambda}})$ and $\hat{Z}_{nk}$ and $\hat{U}_{ngk}$ for all $n, g$ and $k$.    
    \State initial $t=0$ (iteration number);
    \Repeat
    \State Start E-step:
      \State Calculate $\hat{Z}_{nk}^{(t)}$, for all $n$ and $k$, as in (\ref{Z_nk1});
      \State Calculate $\hat{U}_{ngk}^{(t)}$, for all $n$, $g$, and $k$, as in (\ref{U_ng}).
      \State Start M-Step using the $\hat{Z}_{nk}^{(t)}$'s and $\hat{U}_{ngk}^{(t)}$'s:
      \State Compute $\pi_k^{(t+1)}$, for $k=1,\dots,K$, as in (\ref{P_1});
      \State  Compute $\phi_k^{(t+1)}$, for $k=1,\dots,K$, as in (\ref{Phi_1});
      \State Compute $\lambda_{gk}^{(t+1)}$, for $k=1,\dots,K, g=1,\dots,G$, as in (\ref{lambda_1}).
   \Until {$[\ell(\boldsymbol{\theta}^{(t+1)}\,|\,\boldsymbol{y})-\ell(\boldsymbol{\theta}^{(t)}\,|\,\boldsymbol{y})] \leq tol$\, or\, maximum number of iterations is achieved.}
  \end{algorithmic}
\end{algorithm}
\begin{algorithm}[htp]
  \caption{EM algorithm for the ZIP mixture model with covariates}
  \label{alg::ZIP_X}
  \begin{algorithmic}[1]
    \Require
      $\boldsymbol{y}$: matrix of data;
     $\,\boldsymbol{\theta}^{(0)}=\,(\boldsymbol{
\pi}^{(0)},\boldsymbol{\phi}^{(0)},\boldsymbol{\rho}^{(0)},\boldsymbol{\beta}_{0}^{(0)},\boldsymbol{\beta}^{(0)})$: initial parameters;
      $\,tol$: tolerance;
      $\,m$: maximum number of iterations;
      $\,\boldsymbol{x}$: matrix of covariates.
    \Ensure
      optimal set of parameters $\boldsymbol{\hat{\theta}}=(\boldsymbol{\hat{\pi}},\boldsymbol{\hat{\phi}},\boldsymbol{\hat{\beta}}_{0},\boldsymbol{\hat{\rho}},\boldsymbol{\hat{\beta}})$, and $\hat{Z}_{nk}, \hat{U}_{ngk}$ for all $n$, $g$, and $k$.   
    \State initial $t=0$ (iteration number);
    \Repeat
    \State Start E-step:
      \State Calculate $\hat{Z}_{nk}^{(t)}$, for all $n$ and $k$, as in (\ref{Z_nk2});
      \State Calculate $\hat{U}_{ngk}^{(t)}$, for all $n$, $g$, and $k$, as in (\ref{U_ng2}).
      \State Start M-Step using the $\hat{Z}_{nk}^{(t)}$'s and $\hat{U}_{ngk}^{(t)}$'s:
      \State Compute $\pi_k^{(t+1)}$, for $k=1,\ldots,K$, as in (\ref{P_1});
      \State Compute $\phi_k^{(t+1)}$, for $k=1,\ldots,K$, as in (\ref{Phi_1});
      \State Compute $\rho_{gk}^{(t+1)}, \beta_{0g}^{(t+1)}, \beta_{pg}^{(t+1)}$, for all $g$, $k$, and $p$, using the Fisher scoring algorithm.
  
    \Until {$[\ell(\boldsymbol{\theta}^{(t+1)}\,|\,\boldsymbol{y})-\ell(\boldsymbol{\theta}^{(t)}\,|\,\boldsymbol{y})] \leq tol$\, or\, maximum number of iterations is achieved.}
  \end{algorithmic}
\end{algorithm}
\begin{algorithm}[htp]
  \caption{ECM algorithm for the ZINB mixture model without covariates}
  \label{alg::ZINB_NoX}
  \begin{algorithmic}[1]
    \Require
      $y$: Matrix of Data;
     $\boldsymbol{\theta}^{(0)}=\,(\boldsymbol{
\pi}^{(0)},\boldsymbol{\phi}^{(0)},\boldsymbol{\mu}^{(0)},\boldsymbol{\nu}^{(0)})$: initial parameters;
      $tol$: tolerance;
      $m$: maximum number of iterations;
    \Ensure
      optimal set of parameters $\boldsymbol{\hat{\theta}}=(\boldsymbol{\hat{\pi}},\boldsymbol{\hat{\phi}},\boldsymbol{\hat{\mu}},\boldsymbol{\hat{\nu}})$, and $\hat{Z}_{nk}, \hat{U}_{ngk}$, for all $n$, $g$ and $k$.    
    \State initial $t=0$ (iteration number);
    \Repeat
    \State Start E-step;
      \State Calculate $\hat{Z}_{nk}^{(t)}$, for all $n$ and $k$, as in (\ref{Z_nk1_ZINB});
      \State Calculate $\hat{U}_{ngk}^{(t)}$, for all $n$, $g$, and $k$, as in (\ref{U_ng1_ZINB});
      \State Start M-Step using the $\hat{Z}_{nk}^{(t)}$'s and $\hat{U}_{ngk}^{(t)}$'s;
      \State Compute $\pi_k^{(t+1)}$, for $k=1,\ldots,K$, as in (\ref{P_1});
      \State Compute $\phi_k^{(t+1)}$, for $k=1,\ldots,K$, as in (\ref{Phi_1});
      \State Compute $\mu_{gk}^{(t+1)}$, for $k=1,\dots,K$ and $g=1,\dots,G$ as in (\ref{mu_1});
      \State Fix $\mu_{gk}$ at $\mu^{(t+1)}_{gk}$, compute $\alpha_{k}^{(t+1)}$, for $k=1,\ldots,K$, using the Newton--Raphson algorithm via the \textit{theta.ml} function in R;
          
    \Until {$[\ell(\boldsymbol{\theta}^{(t+1)}\,|\,\boldsymbol{y})-\ell(\boldsymbol{\theta}^{(t)}\,|\,\boldsymbol{y})] \leq tol$\, or\, maximum number of iterations is achieved.}
  \end{algorithmic}
\end{algorithm}
\begin{algorithm}[htp]
  \caption{ECM algorithm for the ZINB mixture model with covariates}
  \label{alg::ZINB_X}
  \begin{algorithmic}[1]
    \Require
      $\boldsymbol{y}$: matrix of data;
     $\,\boldsymbol{\theta}^{(0)}=\,(\boldsymbol{
\pi}^{(0)},\boldsymbol{\phi}^{(0)},\boldsymbol{\nu}^{(0)},\boldsymbol{\beta}_{0}^{(0)},\boldsymbol{\rho}^{(0)},\boldsymbol{\beta}^{(0)})$: initial parameters;
        $\, x_{np}$: matrix of covariates;
      $\,tol$: tolerance;
      $\,m$: maximum number of iterations;
    \Ensure
      optimal set of parameters $\boldsymbol{\hat{\theta}}=(\boldsymbol{\hat{\pi}},\boldsymbol{\hat{\phi}},\boldsymbol{\hat{\beta}}_{0},\boldsymbol{\hat{\rho}},\boldsymbol{\hat{\beta}},\boldsymbol{\hat{\nu}})$, and $\hat{Z}_{nk},\,\hat{U}_{ngk}$, for all $n$, $g$ and $k$.    
    \State initial $t=0$ (iteration number);
    \Repeat
    \State Start E-step;
      \State Calculate $\hat{Z}_{nk}^{(t)}$, for all $n$ and $k$, as in (\ref{Z_nk2_ZINB});
      \State Calculate $\hat{U}_{ngk}^{(t)}$, for all $n$, $g$, and $k$, as in (\ref{U_ng2_ZINB});
      \State Start M-Step using the $\hat{Z}_{nk}^{(t)}$'s and $\hat{U}_{ngk}^{(t)}$'s;
      \State Compute $\pi_k^{(t+1)}$, for $k=1,\ldots,K$, as in (\ref{P_1});
      \State Compute $\phi_k^{(t+1)}$, for $k=1,\ldots,K$, as in (\ref{Phi_1});
      \State Fix $\alpha_k$ at $\alpha_{k}^{(t)}$, compute $\beta_{0g}^{(t+1)}$, $\rho_{gk}^{(t+1)}$, and $\beta_{pg}^{(t+1)}$, for all $g$, $k$ and $p$ using the Newton--Raphson algorithm;
      \vspace{0.15cm}
       \State Fix $\beta_{0g}$, $\rho_{gk}$, and $\beta_{pg}$ at $\beta_{0g}^{(t+1)}$, $\rho_{gk}^{(t+1)}$, and $\beta_{pg}^{(t+1)}$, compute $\nu^{(t+1)}$ using the Newton--Raphson algorithm via the \textit{theta.ml} function in R. Let $\alpha_{k}^{(t+1)}= 1/\nu_k^{(t+1)}$.
       \vspace{0.15cm}
   \Until {$[\ell(\boldsymbol{\theta}^{(t+1)}\,|\,\boldsymbol{y})-\ell(\boldsymbol{\theta}^{(t)}\,|\,\boldsymbol{y})] \leq tol$\, or\, maximum number of iterations is achieved.}
  \end{algorithmic}
\end{algorithm}




\clearpage

\bibliography{References}

\end{document}